\documentclass[aps,prd,twocolumn,nofootinbib,superscriptaddress]{revtex4-1}
\usepackage[utf8]{inputenc}
\usepackage{orcidlink}
\usepackage{color}
\usepackage{graphicx}
\usepackage{amsmath}
\usepackage{amssymb}
\usepackage{bm}
\usepackage{acronym}
\usepackage{ifthen}
\usepackage{blindtext}
\usepackage[normalem]{ulem}
\usepackage{hyperref}
\usepackage{etoolbox}
\usepackage{fancyhdr}
\usepackage{xspace}
\usepackage{textcomp}
\usepackage{multirow}
\usepackage[caption=false]{subfig}
\usepackage{lineno}
\usepackage{tabularx}

\def\gw{gravitational wave\xspace}
\def\gwh{gravitational-wave\xspace}
\def\gws{gravitational waves\xspace}

\def\cwh{continuous-wave\xspace}

\newacro{S/N}{signal-to-noise ratio}

\newacro{PN}{post-Newtonian}

\newacro{O3}{the third observing run}

\newacro{O2}{the second observing run}

\newacro{CW}{\emph{Continuous Wave}}

\newcommand{\bea}{\begin{eqnarray}}
\newcommand{\eea}{\end{eqnarray}}
\newcommand{\be}{\begin{equation}}
\newcommand{\ee}{\end{equation}}

\newtoggle{fullauthorlist}
\toggletrue{fullauthorlist}
\newtoggle{endauthorlist}
\toggletrue{endauthorlist}

\begin{document}
\title{All-sky search for gravitational wave emission from scalar boson clouds around spinning black holes in LIGO O3 data}
\date{\today}
\iftoggle{endauthorlist}{
  %
  %
  \let\mymaketitle\maketitle
  \let\myauthor\author
  \let\myaffiliation\affiliation
  \author{The LIGO Scientific Collaboration}
  \author{The Virgo Collaboration}
  \author{The KAGRA Collaboration}
  \email{Full author list given at the end of the article.}
\noaffiliation
}{
\iftoggle{fullauthorlist}{
  \input{KAGRA-LSC-Virgo-Authors-Aug-2020-aas}
}
{
  \author{The LIGO Scientific Collaboration}
  \affiliation{LSC}
  \author{The Virgo Collaboration}
  \affiliation{Virgo}
  \author{The KAGRA Collaboration}
  \affiliation{KAGRA}
  
}
}

\begin{abstract}
This paper describes the first all-sky search for long-duration, quasi-monochromatic \gwh signals emitted by ultralight scalar boson clouds around spinning black holes using data from the third observing run of Advanced LIGO. We analyze the frequency range from 20~Hz to 610~Hz, over a small frequency derivative range around zero, and use multiple frequency resolutions to be robust towards possible signal frequency wanderings. Outliers from this search are followed up using two different methods, one more suitable for nearly monochromatic signals, and the other more robust towards frequency fluctuations.
We do not find any evidence for such signals and set upper limits on the signal strain amplitude, the most stringent being $\approx10^{-25}$ at around 130~Hz. 
We interpret these upper limits as both an ``exclusion region'' in the boson mass/black hole mass plane and the maximum detectable distance for a given boson mass, based on an assumption of the age of the black hole/boson cloud system.
\end{abstract}
        
        \maketitle

\section{Introduction}\label{sec:intro}

Theories of beyond standard model physics allow for the production of ultralight bosons that could constitute a portion or all of dark matter \cite{Preskill:1982cy,Abbott:1982af,Dine:1982ah,Arvanitaki:2009fg,Arvanitaki:2010sy,arvanitaki2015discovering}. If such ultralight bosons exist, they could appear around rotating black holes due to quantum fluctuations \cite{Brito:2015oca}. They would then scatter off and extract angular momentum from these black holes, and form macroscopic clouds, i.e. boson condensates, through a superradiance process \cite{Brito:2015oca,Arvanitaki:2016qwi}. The energy structure of the clouds resembles that of a hydrogen atom, earning boson clouds the name ``gravitational atoms'' in the sky. After the black hole spin decreases below a threshold, the superradiance process stops and the cloud depletes over time mainly through annihilation of bosons into gravitons, which generates quasi-monochromatic, long-duration \gws when the self-interaction for bosons is weak.

The signal would also have a small, positive frequency variation, known as a ``spin-up'', that would arise from the classical contraction of the cloud over time as it loses mass \cite{Brito:2015oca}. The values of the spin-up depend on whether we consider scalar \cite{Brito:2017zvb}, vector \cite{baryakhtar2017black,siemonsen2020gravitational}, or tensor \cite{Brito:2020lup} boson clouds, but in this search, we consider scalar boson cloud signals with very small spin-ups, of maximum $\mathcal{O}(10^{-12})$ Hz/s.
Recently, the effect on \gwh emission from boson self-interactions has been studied for the scalar case \cite{2021PhRvD.103i5019B}, showing that the signal can be significantly affected, including the magnitude of the spin-up, as the self-interaction increases.   

The \gwh signal frequency depends primarily on the mass of the boson, and weakly on the mass and spin of the black hole. Recently, exclusion regions of the black hole / scalar boson mass parameter space were calculated using upper limits from an all-sky search for spinning neutron stars in LIGO/Virgo data from the second observing run (O2)~\cite{palomba2019direct}. Furthermore, a search for boson clouds from Cygnus X-1 was performed with a hidden Markov model (HMM) method \cite{isi2019directed} on the same data set, and disfavored particular boson masses for this object \cite{Sun:2019mqb}. It has been suggested \cite{PhysRevD.102.063020} that current methods for all-sky searches, such as the one used in \cite{palomba2019direct}, could suffer a significant sensitivity penalty in the extreme case of a population of $10^8$ black holes in the Milky Way, due to the superposition of many signals in a small frequency range (as a consequence of the signal's frequency dependence at leading order on the boson mass). A detailed study \cite{pierini2021inprep} has quantified this effect, showing that the actual average loss of sensitivity is at most of about 15$\%$ for a signal ``density'' of 1-3 signals per frequency bin, while it rapidly reduces, and becomes negligible, for both lower and higher densities.   

In addition to quasi-monochromatic \gws emitted by boson clouds around isolated black holes, it is also possible to probe the existence of boson clouds in binary black hole coalescences. One avenue requires measurements of spin: in principle, boson clouds will extract mass and spin from black holes, resulting in low spin values for black holes compared to those in a universe without boson clouds; in practice, individual black hole spins are hard to measure \cite{Vitale:2014mka,Purrer:2017uch}, and the current spin distribution of black holes depends on both the mass of boson clouds and the distribution of spins when the black holes were born \cite{Belczynski:2017gds,Gerosa:2018wbw}.  {It is therefore interesting to combine spin measurements from various black hole mergers, via hierarchical Bayesian inference, to obtain constraints on boson cloud/spin interactions \cite{Ng:2019jsx,Ng:2020ruv}. }
Additionally, the presence of a boson cloud will induce a change in the waveforms used in compact binary searches, e.g. new multipole moments and tidal effects (parameterized as Love numbers) \cite{baumann2019probing,Yang:2017lpm}, and such differences in binary black hole mergers may actually be detectable depending on the boson mass and field strength \cite{Choudhary:2020pxy}.
It has also been proposed to combine multi-band observations of black hole mergers and boson cloud signals when future detectors are online \cite{Ng:2020jqd}, and to search for a stochastic \gwh background from scalar and vector boson cloud signals \cite{Brito:2017wnc,Tsukada:2018mbp,Tsukada:2020lgt,Yuan:2021ebu}. Complementary methods \cite{pierce2019dark,Miller:2020vsl} and searches \cite{guo2019searching,LIGOScientific:2021odm} for ultralight scalar and vector bosons have also been developed that use the \gwh interferometers as particle detectors, which provide additional constraints on the boson mass.
Ultralight bosons can therefore have different effects on different types of \gwh signals.

This paper presents results from the first all-sky search tailored to \gws from depleting scalar boson clouds around black holes using the Advanced LIGO~\citep{2015CQGra..32g4001L} data in the third observing run (O3). Although the expected \gwh signal is almost monochromatic, it could come from anywhere in the sky; thus, the Doppler modulations from the relative motion of the Earth and the source, of $\mathcal{O}(10^{-4}f)$, where $f$ is the frequency of the signal, are a priori unknown. A fully coherent search, in which we take a single Fourier Transform of the whole data set and look for peaks in the power spectrum after demodulating the data, is therefore impossible to perform in each sky direction because of limited computational power. Indeed, the number of sky positions to search over scales with the square of both the Fourier Transform length and the frequency, and would amount to $\mathcal{O}(10^{20})$ at high frequencies \cite{Astone:2014esa}.  This means that we must employ semi-coherent methods, in which we break the data into smaller chunks in time that are analyzed coherently, and then combined incoherently,
to keep the computational cost under control while retaining the desired sensitivity \cite{riles2017recent,sieniawska2019continuous,Tenorio:2021wmz}. 
Moreover, we adopt a multi-resolution approach, equivalent to considering several different Fourier Transform lengths \cite{d2018semicoherent}, in order to be sensitive towards possible unpredicted frequency fluctuations of the \gwh signal.

This paper is organized as follows: in section \ref{sec:model}, we describe our model for a \gwh signal resulting from a depleting scalar boson cloud around a rotating black hole; in section \ref{sec:method}, we explain our method to search for such signals; in section \ref{sec:data}, we provide information on the data sets we use in our analysis; in section \ref{sec:results}, we present the results of the search and our upper limits, including their astrophysical interpretations in terms of constraints in the boson/black hole mass plane and on the maximum detectable distance; finally, in section \ref{sec:conclusions}, we discuss prospects for future work. Appendices contain details on the follow-up of selected outliers. 

\section{The signal} \label{sec:model}

An isolated black hole is born with a defining mass, spin and charge \cite{thorne2000gravitation}. Ultralight boson particles around a black hole will scatter off it and extract energy and momentum from it if the so-called \emph{superradiant} condition is satisfied, i.e. $\omega<m\Omega$ \cite{Bekenstein:1973mi,Brito:2015oca} where $\omega$ is the angular frequency of the boson (related to the boson mass linearly), $m$ is the azimuthal quantum number in the hydrogen-like gravitational atom, and $\Omega$ is the angular frequency of the rotating black hole's outer horizon. Because the bosons have a finite mass, they become gravitationally bound to the black hole, which allows for successive scatterings and a build-up of a macroscopic cloud, extracting up to about $10\%$ of the black hole mass \cite{Brito:2015oca}. This process is maximized when the particle's reduced Compton wavelength is comparable to the black hole radius: 
\begin{equation}
\frac{\hbar c}{m_b}\simeq \frac{GM_{\rm BH}}{c^2},
\end{equation}
where $m_b$ is the boson mass-energy, $M_{\rm BH}$ is the black hole mass, $\hbar$ is the reduced Planck's constant, and $c$ is the speed of light. The typical time scale of the superradiance phase is \cite{Brito:2017zvb}
\begin{equation}
    \tau_{\rm inst} \approx 27\left(\frac{M_\mathrm{BH}}{10M_{\odot}}\right)\left(\frac{\alpha}{0.1}\right)^{-9}\left(\frac{1}{\chi_i}\right)~\mathrm{days},
\end{equation}
where $\chi_i$ is the black hole initial dimensionless spin, and
\begin{equation}
    \alpha = \frac{G M_{\rm BH}}{c^3} \frac{m_b}{\hbar}, \label{eq:alpha} 
\end{equation}
is the fine-structure constant in the gravitational atom. 

Once the superradiant condition is no longer satisfied, the growth stops, and the cloud begins to give off energy primarily via annihilation of particles in the form of \gws \cite{isi2019directed}, over a typically much longer time scale which, for $m=1$ and in the limit $\alpha \ll 0.1$, is given by
\begin{equation}
\tau_{\rm gw} \approx 6.5 \times 10^4\left(\frac{M_\mathrm{BH}}{10M_{\odot}}\right)\left(\frac{\alpha}{0.1}\right)^{-15}\left(\frac{1}{\chi_i}\right)~\mathrm{years}.
\label{eq:taugw}
\end{equation}
For scalar bosons, these clouds have a much shorter growth time scale compared to the time scale for them to deplete \cite{Brito:2017zvb}; thus, if they exist now, they are more likely to be depleting rather than growing. The \gwh emission is at a frequency $f_\text{gw}$ \cite{palomba2019direct}:
\begin{eqnarray}
f_\text{gw} & \simeq & 483\,{\rm Hz} \left(\frac{m_\text{b}}{10^{-12}~\text{eV}}\right) \nonumber\\
&& \times \left[1-7\times 10^{-4}\left(\frac{M_\text{BH}}{10M_{\odot}}\frac{m_\text{b}}{10^{-12}~\text{eV}}\right)^2\right].
\label{eq:fgw}
\end{eqnarray}
In fact, this frequency slowly increases in time due to the cloud self-gravity and particle self-interaction, quantified by the boson self-interaction constant $F_b$ \cite{2021PhRvD.103i5019B}. As in \cite{2021PhRvD.103i5019B}, we consider a scalar boson that, in addition to a non-zero mass, has a quartic interaction with an extremely small coupling $\lambda$, and characterize this coupling in terms of the parameter $F_b:=m_b/\sqrt{\lambda}$. In the context of an axion-like particle, $F_b$ would roughly correspond to the symmetry breaking energy scale.

Specifically, if $F_b \gtrsim 2 \times 10^{18}$~GeV, the spin-up is dominated by the cloud self-gravity and is given by
\begin{equation}
\label{eq:fanni}
\dot{f}_\mathrm{gw}\approx 7\times 10^{-15} \left(\frac{m_\text{b}}{10^{-12}~\text{eV}}\right)^2\left(\frac{\alpha}{0.1}\right)^{17}~\mathrm{Hz/s}.
\end{equation}
With this condition on $F_b$, our search probes energies at the Planck scale, and is sensitive to the QCD axion with a mass of $\sim 10^{-13}$--$10^{-12}$ eV \cite{2021PhRvD.103i5019B}.
However, for smaller $F_b$ a spin-up term, due to the cloud self-gravity and change of the black hole mass, of the form 
\begin{equation}
\label{eq:fdot_tr}
\dot{f}\approx 10^{-10} \left(\frac{m_\text{b}}{10^{-12}~\text{eV}}\right)^2\left(\frac{\alpha}{0.1}\right)^{17}\left(\frac{10^{17}\mathrm{GeV}}{F_b}\right)^4~\mathrm{Hz/s},
\end{equation}
is dominant in the intermediate self-interaction regime, for $8.5 \times 10^{16}(\alpha/0.1)~\mathrm{GeV} \lesssim F_b \lesssim 2 \times 10^{18}~\mathrm{GeV}$ (a term with a similar scaling appears due to energy level transitions). In this regime, the signal duration significantly shortens, thus reducing the chance of detection. In the strong self-interaction regime, when $F_b \lesssim 8.5 \times 10^{16}(\alpha/0.1)~\mathrm{GeV}$, the spin-up can be parametrized as
\begin{equation}
\label{eq:fdot_se}
\dot{f}\approx 10^{-10} \left(\frac{m_\text{b}}{10^{-12}~\text{eV}}\right)^2\left(\frac{\alpha}{0.1}\right)^{17}\left(\frac{10^{17}\mathrm{GeV}}{F_b}\right)^6~\mathrm{Hz/s},
\end{equation}
and the signal strength rapidly decreases with increasing interaction, making the detection of annihilation signals basically impossible for current detectors.
In Sec.~\ref{sec:astro}, we will briefly discuss the implication of our search results in relation to the value of $F_b$. {Moreover, at the analysis level we do not want to exclude the possibility that the emitted signal frequency evolves in a more complicated manner than we model, e.g. there are random fluctuations. In Sec. \ref{sec:peakmav}, we describe a computationally cheap method to deal with such frequency variations.}

The signal we are considering have lower amplitudes than those from compact binary coalescences \cite{PhysRevD.102.063020}, with initial values (for $\alpha \ll 0.1$)
\begin{equation} 
\label{eq:h0}
h_0 \approx 6 \times 10^{-24} \left(\frac{M_\mathrm{BH}}{10 M_\odot}\right)
\left(\frac{\alpha}{0.1}\right)^7 \left(\frac{1\,\rm kpc}{D}\right)
\left({\chi_i - \chi_c}\right),
\end{equation}
where 
\begin{eqnarray}
\chi_c \approx \frac{4 \alpha}{4\alpha^2 + 1},
\end{eqnarray}
is the black hole spin when the superradiance is saturated, and $D$ is the source distance. 
The emitted signal amplitude decreases in time as the clouds are depleted, according to
\begin{equation} 
h(t) = \frac{h_0}{1+\frac{t}{\tau_{\rm gw}}}.
\label{eq:hoft}
\end{equation}

The equations above are analytic approximations to the true behavior of a \gw originating from boson clouds. When we consider $\alpha \sim 0.1$, we must account for the difference between the numerical and analytic solutions, which is $\sim 3$ in energy at the highest $\alpha$ considered here, $\alpha\sim0.15$  We obtain this factor of $\sim 3$ comparing the black and red curves in Fig. 2 of \cite{Brito:2017zvb} at $\alpha\sim 0.15$.

\section{Search method} \label{sec:method}
The semi-coherent method we employ is based on the Band Sampled Data (BSD) framework \cite{piccinibsd}, which is being used in an increasing number of continuous-wave searches, due to its flexibility and computational efficiency \cite{piccinni2020directed,abbott2021snr,abbott2020gravitational}. BSD files are data structures that store a reduced analytic strain-calibrated time series in 10-Hz/one-month chunks. To construct BSD files, we take a Fourier transform of a chunk of time-series strain data, extract a 10-Hz band, keep only
the positive frequency components, and inverse Fourier
Transform to obtain the reduced analytic signal.

\begin{figure}[htb]
    \centering
    \includegraphics[width=\columnwidth]{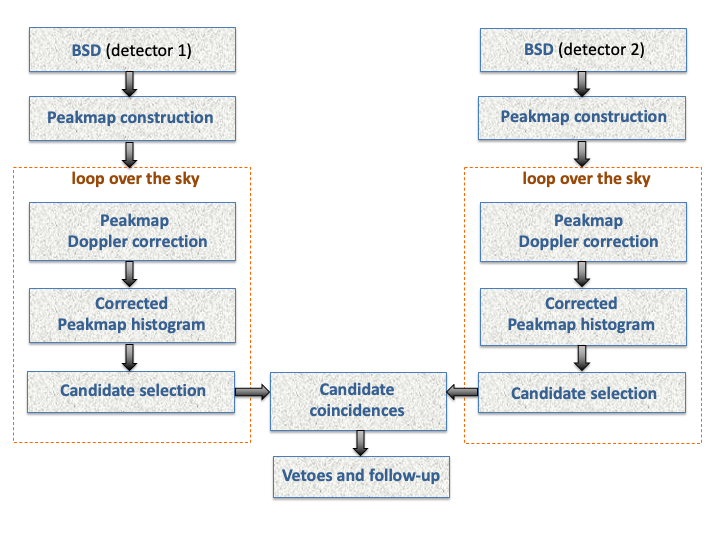}
    \caption{Scheme of the search pipeline.} 
    \label{fig:pipescheme}
\end{figure}

The main steps of the analysis pipeline are schematically shown in Fig. \ref{fig:pipescheme} and are summarized below.

\subsection{Peakmap construction} \label{subsec:peakmap}
\begin{figure}[htb]
    \centering
    \includegraphics[width=\columnwidth]{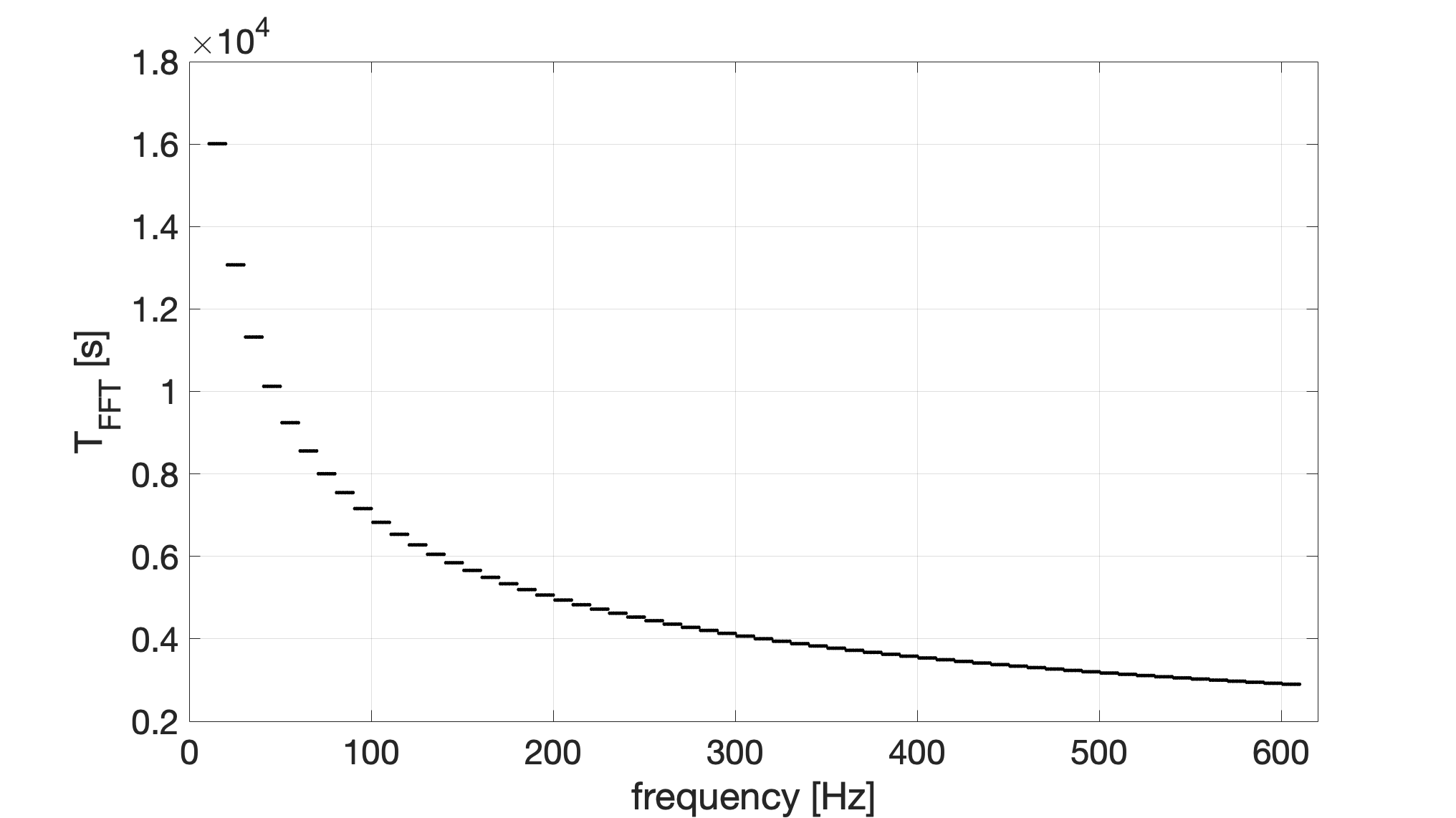}
    \caption{Chunk duration $T_{\rm FFT}$ as a function of frequency. The $T_{\rm FFT}$ is fixed within each 10-Hz band.}
    \label{fig:tfft}
\end{figure}

Our analysis starts with a set of BSD files covering the whole third observing run and frequencies between 20 Hz and 610 Hz. The maximum frequency is chosen to ensure the computational cost of the analysis is reasonable, and is consistent with the fact that the most relevant part of the accessible parameter space of the black hole-boson system corresponds to the intermediate frequencies. Each of these time series is divided into chunks of duration
\begin{equation}
T_{\rm FFT}=\frac{1}{\Omega_{\rm rot}}\sqrt{\frac{c}{2f_{\rm 10}R_{\rm rot}}},
\label{eq:tfft}
\end{equation}
where $f_{\rm 10}$ is the maximum frequency of each corresponding 10-Hz band, $\Omega_{\rm rot}=2\pi/86164.09~$rad/s is the Earth sidereal angular frequency, and $R_{\rm rot}$ is the Earth rotational radius, which we conservatively take as that corresponding to the detector at the lowest latitude ($R_{\rm rot} = 5.5 \times 10^6$~km for the Livingston detector). This $T_{\rm FFT}$ length guarantees that, if we take the Fourier Transform of the chunk, the power spread of any possible \cwh signal, caused by the motion of the Earth with respect to the source, is fully contained in one frequency bin with a width of $\delta f=1/T_{\rm FFT}$ in each chunk. In Fig.~\ref{fig:tfft}, the chunk duration is shown as a function of the frequency.
For each of these chunks, with at least a $50\%$ of non-zero values, the square modulus of the Fourier Transform is computed (by means of the FFT algorithm) and divided by an estimation of the average spectrum over the same time window. The resulting quantity has a mean value of one independently of the frequency - for this reason it is called the {\it equalized} spectrum - and takes values significantly different from one when narrow frequency features are present in the data. We select prominent peaks in the equalized spectra, defined as time-frequency pairs that correspond to local maxima and have a magnitude above a threshold of $\theta = 2.5$ \cite{Astone:2014esa}. The collection of the these time-frequency peaks forms the so-called {\it peakmap}, see e.g. \citep{Astone:2014esa} for more details. 

\subsection{Doppler effect correction}
We build a suitable grid in the sky such that, after the Doppler effect correction for a given sky direction, any residual Doppler effect always produces an error in frequency within half a frequency bin \cite{Astone:2014esa}. For each sky direction in the grid, the peaks in the peakmap are shifted in order to compensate for the time-dependent Doppler modulation, according to the relation
\begin{equation}
    f_{0,\rm{t_k}}=\frac{f_{\rm obs,t_k}}{1+\frac{\vec{v}_{\rm t_k} \cdot \hat{n}}{c}},
    \label{eq:doppcorr}
\end{equation}
where $f_{\rm obs,t_k}$ is an observed frequency peak at the time $t_k$, $\vec{v}_{\rm t_k}$ is the detector's velocity, and $\hat{n}$ is the unit vector identifying the sky direction. 

\begin{figure}[htb]
    \centering
    \includegraphics[width=\columnwidth]{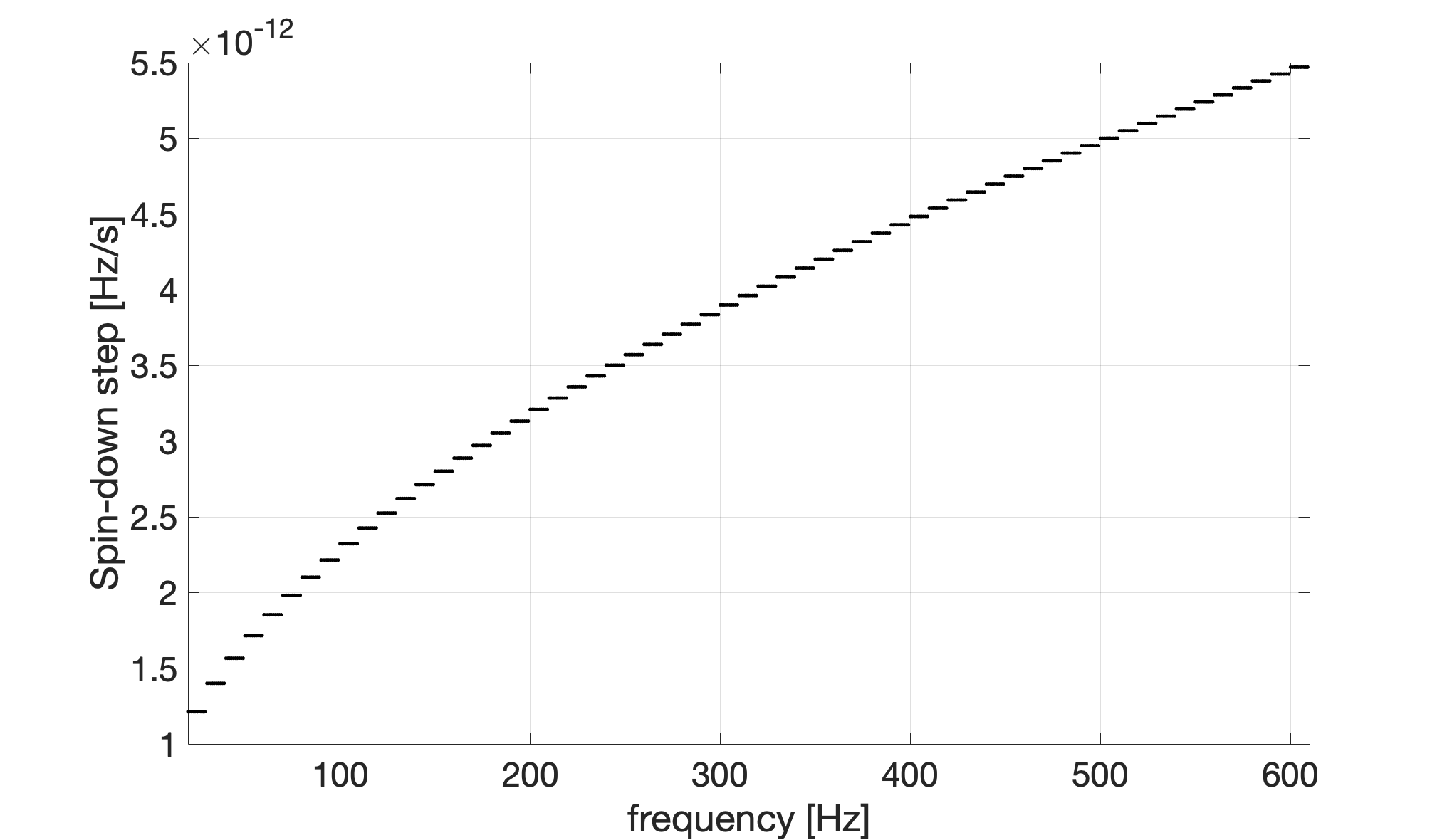}
    \caption{Bin size of the frequency time derivative, $\delta \dot{f}$, a function of the frequency, given the full observing time of O3 and $T_\mathrm{FFT}$ chosen for each 10-Hz band. The search by default covers a frequency time derivative range of $\dot{f} \in [-\delta \dot{f}/2, \delta \dot{f}/2]$. Such a range contributes to set constraints on the portion of the boson mass/black hole mass explored in the search.}
    \label{fig:spindownstep}
\end{figure}

As we have described in Sec.~\ref{sec:model}, the signal has an intrinsic spin-up, associated with the cloud depletion. Although we do not apply an explicit correction for the spin-up, the analysis retains most of the signal power - with a maximum sensitivity loss less than a few percent\footnote{In principle, the maximum sensitivity loss would be of about 36$\%$, when the corrected signal frequency falls in the middle among two consecutive frequency bins. In practice, however, the application of the moving average on the peakmaps, see section \ref{sec:peakmav}, allows us to recover part of the lost signal power, reducing the maximum loss to about 19$\%$ (when applying a moving average of two bins, i.e. a window with W=2).} - as long as the corresponding frequency variation during the full observation time $T_\mathrm{obs}$ is confined within $\pm 1/2$ bin of the frequency evolution rate $\delta \dot{f}$, given by
\begin{equation}
    \delta \dot{f} = \frac{1}{2T_\mathrm{obs} T_\mathrm{FFT}}.
\end{equation}
The $\delta \dot{f}$ value is plotted as a function of the frequency in Fig.~\ref{fig:spindownstep}.

\subsection{Peakmap projection}\label{subsec:peakmap}
Once we have applied the Doppler correction, we project the peakmap onto the frequency axis and select the most significant frequencies for further analysis. Indeed, if a monochromatic signal comes from a given direction, we expect its associated peaks to be aligned in frequency and appear as a prominent peak in the projected peakmap, when the signal is strong enough. The statistics of the peakmap are described in detail in \citep{Astone:2014esa}, so here we only provide a brief review. In the absence of signals, the probability $p_0$ of selecting a peak depends on the noise distribution. In the case of Gaussian noise, it is given by 
\begin{equation}
    p_0=e^{-\theta}-e^{-2\theta}+\frac{1}{3}e^{-3\theta},
\end{equation}
where $\theta=2.5$ is the threshold we apply to construct the peakmap. 
On the other hand, if a signal with spectral amplitude (in units of equalized noise)
\begin{equation}
    \lambda=\frac{4|\tilde{h}(f)|^2}{T_{\rm FFT}S_n(f)},
\end{equation}
where $\tilde{h}(f)$ is the signal Fourier Transform, and $S_n(f)$ is the detector noise power spectral density, is present, the corresponding probability $p_{\lambda}$ of selecting a signal peak, for weak signals with respect to the noise level, is given by
\begin{equation}
    p_{\lambda}=p_0 + \frac{\lambda}{2}\theta \left(e^{-\theta}-2e^{-2\theta}+e^{-3\theta}\right).
\end{equation}
The statistic of the peakmap projection is, for well-behaved noise, a binomial with expectation value $\mu=N p_0$, where $N$ is the number of FFTs, and standard deviation $\sigma=\sqrt{N p_0 (1-p_0)}$. In the presence of a signal, the expected number of peaks at the signal frequency (after the Doppler correction) is $N p_{\lambda}$. It is then useful to introduce the Critical Ratio CR:
\begin{equation}
    \rm{CR}=\frac{n-\mu}{\sigma},
\end{equation}
where $n$ is the actual number of peaks in a given frequency bin,
which, in the limit of large N (in practice, greater than a $\sim30$), closely follows a Gaussian distribution with zero mean and unity standard deviation.

\subsection{Peakmap moving averages}
\label{sec:peakmav}

The signal emitted by a boson cloud is not exactly monochromatic, due to the presence of a spin-up, see Sec. \ref{sec:model}. {Moreover, further unpredicted frequency fluctuations could be present, due to our limited knowledge of the physical processes actually taking place. A signal with varying frequency can be characterized by its \textit{coherence time}, defined as the minimum amount of time over which the signal frequency variation exceeds the search frequency bin.} 
For the sake of robustness, in order to deal with {spin-up values which cause a frequency variation larger than half a frequency bin during the observing time $T_{\rm obs}$, see Sec. \ref{subsec:peakmap}, and - especially - to take into account} possible uncertainties in the signal morphology, we apply a sequence of moving averages (MA) over a window width $W$ varying from two to ten frequency bins, starting from the Doppler corrected peakmap projection. This procedure can be shown to be equivalent -- and computationally much more efficient -- to build peakmap projections with an equivalent FFT duration $\approx T_{\rm FFT}/W$. While these moving averages reduce the sensitivity to signals with a characteristic coherence time larger than $T_{\rm FFT}$, e.g. purely monochromatic 
signals, they provide better sensitivity to signals with a typical coherence time comparable to the equivalent FFT duration \citep{d2018semicoherent}. {At the same time, they allow to partially recover sensitivity to spin-ups larger than those shown in Fig. \ref{fig:spindownstep}.}

\subsection{Candidate selection and coincidences}
\label{sec:candidates}
We use the CR as a detection statistic to identify potentially interesting outliers. Outliers are uniformly selected in the parameter space by taking the two points with the highest CR in each peakmap projection, for each sky position in every 0.05-Hz frequency band, independent of the actual CR values, as long as it is above a minimum value of 3.8. 

In the next step, we identify coincident pairs among outliers found in the Hanford and Livingston data, corresponding to the same $W$, by using a dimensionless coincidence window of 3. In other words, two outliers, one from Hanford and the other from Livingston, are considered coincident if the dimensionless distance
\begin{equation}
    d = \sqrt{\left(\frac{\Delta f}{\delta f}\right)^2+\left(\frac{\Delta \lambda}{\delta \lambda}\right)^2+ \left(\frac{\Delta \beta}{\delta \beta}\right)^2}\le 3
\end{equation}
where $\Delta f,~\Delta \lambda,~\Delta \beta$ are the differences between parameter values of the outliers found in Hanford (H1) and Livingston (L1), and $\delta f,~\delta \lambda,~\delta \beta$ are the corresponding bin sizes\footnote{The sky position bin sizes in general are different for two outliers identified at different sky positions, so the average among the corresponding bin sizes is actually used.}. The coincidence window of 3 is chosen according to the studies carried out using simulated signals and is consistent with the choice in standard all-sky searches for neutron stars \cite{pisarski2019all,abbott2017all}. 

\subsection{Post-processing}
\label{sec:postproc}
Coincident outliers are subject to a sequence of post-processing steps in order to discard those incompatible with an astrophysical signal. 

The first veto is to check if some of the outliers are compatible with known narrow-band instrumental disturbances, also known as {\it noise lines}. An outlier is considered as caused by a noise line, and thus discarded, if its Doppler band, defined by $\Delta f_{\rm dopp}=\pm  10^{-4}f$, where $f$ is the outlier frequency, intersects the noise line.

The next is a consistency veto, in which a pair of coincident outliers are discarded if their CR is not compatible with the detectors' noise level at the corresponding frequency. Specifically, we veto those coincident outliers for which the CR in one detector is more than a factor of 5 different than that in the other detector, after being weighted by the detector's power spectral density \cite{abbott2017all,palomba2019direct}. This choice, also used in standard all-sky searches for continuous waves from neutron stars, is motivated by the desire to safely eliminate an outlier only if the discrepancy is really significant and the CR is large in at least one of the two detectors (i.e., a real astrophysical signal is not expected to behave that way).

The previous steps are all applied to coincident outliers separately for each value of the MA window width $W$. In case of outliers coincident across different $W$ values, only the one with the highest CR is kept\footnote{This restriction reduces the computational cost with only a small loss in sensitivity.}.

\subsection{Follow-up}
Outliers with $W=1$, the case in which we do not apply moving averages, could be due to monochromatic signals or signals with very small fluctuations in frequency, while outliers with $W>1$ are more compatible with signals characterized by a larger degree of frequency fluctuation. These two sets of candidates are then followed up with different methods, one based on the Frequency-Hough algorithm, and used e.g. in \cite{pisarski2019all}, and the other using the Viterbi method, see e.g. \cite{Sun:2019mqb}. Both methods are briefly explained in Appendix \ref{sec:fumethods}.

\section{Data}\label{sec:data}

We use data collected by the Advanced LIGO \gwh detectors over the O3 run, which 
lasted from April 1st, 2019 to March 27th, 2020, with a one-month break in data collection in October 2019. The duty cycles of the two detectors, H1 and L1, are $\sim 76\%$ and $\sim 77\%$, respectively, during O3.

In the case of a detection, the calibration uncertainties in the detector strain data stream could impact the estimates of the boson properties. Without a detection, these uncertainties could affect the estimated instrument sensitivity and inferred upper limits on astrophysical source properties. 
The analysis uses the ``C01'' version of calibrated data\footnote{{This is equivalent to the \href{https://www.gw-openscience.org/O3/}{publicly-available} frame data titled ``HOFT\_CLEAN\_SUB60HZ\_C01\_AR''}}, 
which has estimated maximum uncertainties (68\% confidence interval) of $\sim 7\%$/$\sim 11 \%$ in magnitude and $\sim 4 \deg$/$\sim 9 \deg$ in phase, over the first/second halves of O3~\citep{Sun:2020wke,Sun:calibrationO3b}. 
The frequency-dependent uncertainties vary over the course of a run but do not change by large values. The time-dependent variations could lead to errors in opposite directions at a given frequency during different time periods. In addition, the calibration errors and uncertainties are different at the two LIGO sites. By integrating the data over the whole run, we expect that the impact from the time-dependent, frequency-dependent calibration uncertainties cancels out to some extent, and the overall impact on the inferred upper limits is within a level of $\sim 2\%$. Thus we do not explicitly consider the calibration uncertainties in our analysis.

Due to the presence of a large number of transient noise artifacts, a \emph{gating} procedure \cite{gatingDocument,y_stochasticGatingDocument} has also been applied to the LIGO data. This procedure applies an inverse Tukey window to the LIGO data at times when the root-mean-square value of the whitened strain channel in the band of 25--50~Hz or 70--110~Hz exceeds a certain threshold. Only $0.4\%$ and $\sim 1\%$ of data is removed for Hanford and Livingston, respectively, and the improvement in data quality from applying such gating is significant, as seen in the stochastic and continuous gravitational-wave analyses in O3 \cite{Abbott:2021xxi}.

\section{Results}\label{sec:results}
In this section, we present the results of the analysis described in Sec.~\ref{sec:method}. With no candidates surviving the follow-up, we present the upper limits on the signal strain amplitude in Sec.~\ref{sec:upperlimits}, and the astrophysical implications in Sec.~\ref{sec:astro}.

Coincident outliers produced by the main search are
first pruned in order to reduce their number to a manageable level.
A further reduction is based on a close case-by-case inspection of the strongest outliers, as described in the following.  

\subsection{Outlier vetos}

First, we remove outliers due to {\it hardware injections}, which simulate the gravitational wave signals expected from spinning neutron stars, added for testing our analysis methods by physically moving the mirrors as if a signal had arrived. See Appendix~\ref{sec:hi} for more details.  

\begin{figure*}[htb]
    \centering
    \includegraphics[width=\columnwidth]{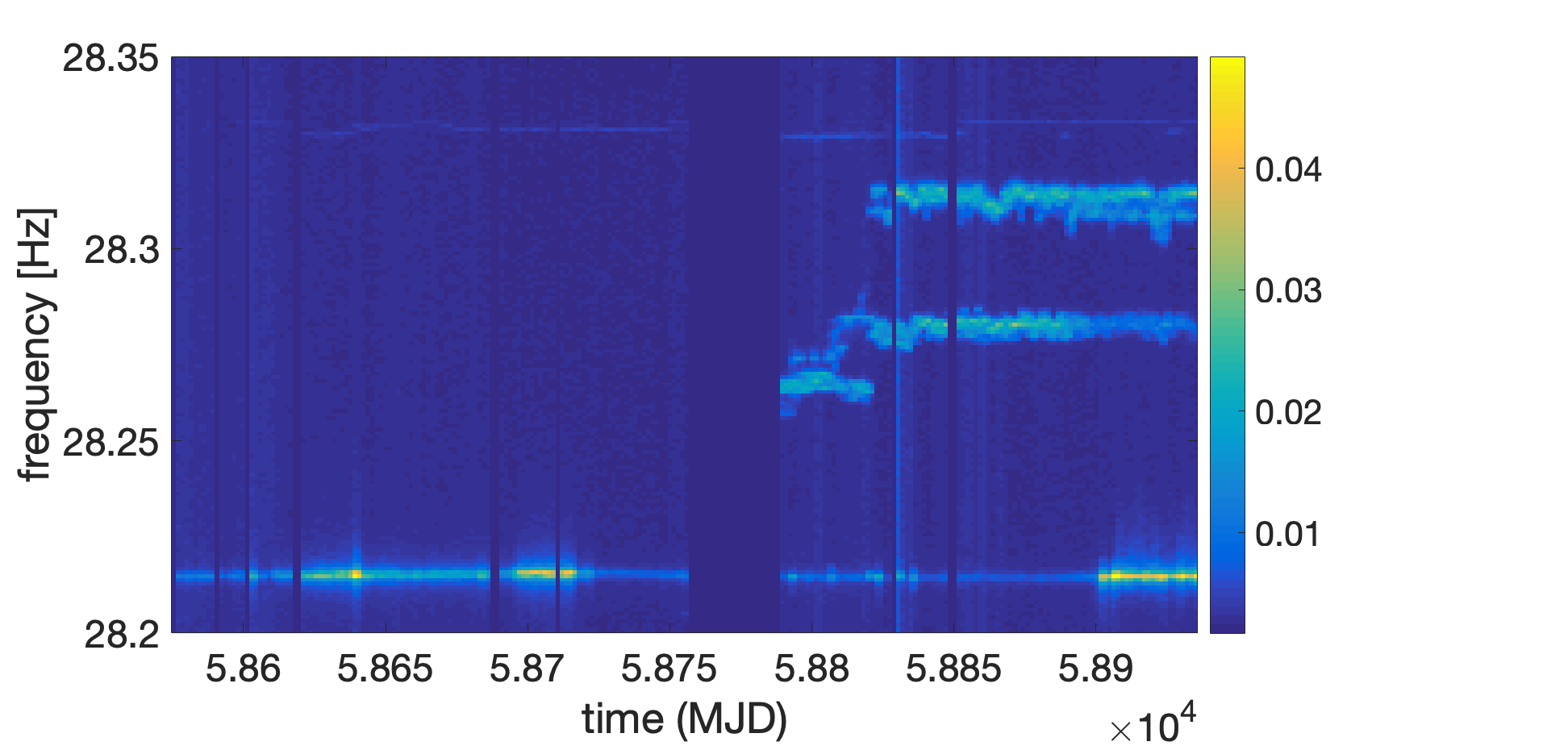}
    \includegraphics[width=\columnwidth]{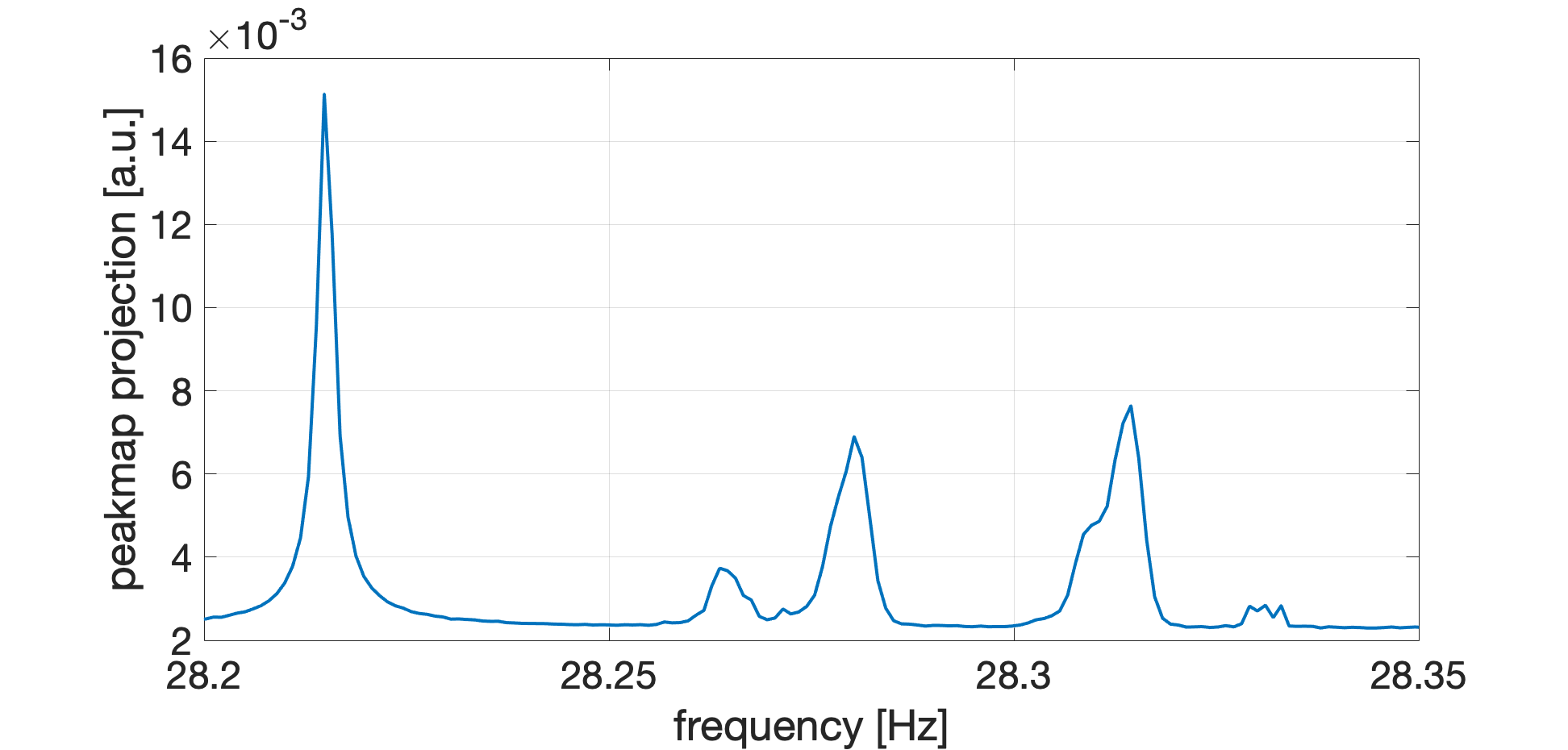}
    \caption{Time-frequency peakmap (left) and corresponding projection (right) in the frequency band of 28.2--28.35 Hz over the full run. Several instrumental artifacts are visible, e.g., strong lines with varying amplitudes ($\sim 28.22$~Hz), weaker lines with sudden frequency variation ($\sim 28.33$~Hz), and broad transient disturbances (vertical bright lines). In particular, the features around 28.28~Hz in the second half of O3 are responsible for a big bunch of outliers with frequency in the range of 28.279--28.283~Hz.}
    \label{fig:disturbance_28}
\end{figure*}

Moreover, we find several bunches of outliers which are not compatible with astrophysical signals, and thus are also discarded and not followed up. In these cases, the time-frequency peakmaps and spectra show the presence of broadband disturbances or lines, 
Examples of such disturbances are shown in Fig.~\ref{fig:disturbance_28}, where we plot on the left the time-frequency peakmap in the frequency range of 28.2--28.35~Hz over the whole run and on the right the corresponding projection onto the frequency axis. A wide and strong transient disturbance responsible for many outliers with frequency in the range of 28.279--28.283~Hz is clearly visible. 

In addition, we find an excess of outliers at several multiples of 10~Hz in both H1 and L1 data, starting from 230~Hz. These are artifacts produced by the procedure to build BSDs\footnote{The artifacts are a well-known consequence of the 10 Hz baseline used to build BSD files. They could be avoided by constructing the BSDs with interlaced frequency bands.}. As it is extremely unlikely that the real astrophysical signals are coincident with multiples of 10~Hz, we veto all candidates within a range of $\pm 10^{-4}$~Hz around each multiple of 10~Hz.  

Finally, we apply a final threshold on the CR in order to further reduce the number of outliers to a small enough number to permit follow up. In particular, we use different thresholds ${\rm CR}_{\rm thr}$ = 4.1 and 4.4 for outliers with $W=1$ and $W>1$, respectively. 
This procedure leaves us with 9449 outliers with $W=1$ and 7816 outliers with $W > 1$. The choice of threshold does not affect the upper limits presented later, since those are based on the outliers with the highest CR.

Such outliers are followed-up with two different methods: a method based on the Frequency-Hough algorithm if the outliers are characterized by $W=1$ (Appendix~\ref{sec:fu_FH}), and a method based on a HMM tracking scheme (the Viterbi algorithm) if the outliers are characterized by $W>1$ (Appendix~\ref{sec:fu_viterbi}). No potential CW candidate remains after the follow-up. Details of the follow-up procedure and results are presented in Appendix~\ref{app:cand}.

\subsection{Upper limits}
\label{sec:upperlimits}
Having concluded that none of the outliers is compatible with an astrophysical signal, we compute $95\%$ confidence level (C.L.) upper limits placed on the signal strain amplitude. 

The limits are computed, for every 1-Hz sub-band in the range of 20--610~Hz, with the analytic formula of the Frequency-Hough sensitivity, given by Eqn.~(67) in \cite{Astone:2014esa}, where the relevant quantities, i.e. an estimate of the noise power spectral density $S_n$ and the CR, are obtained from the O3 data used in this search. Specifically, for each 1-Hz subband, we take the maximum CR of the outliers, separately for Hanford and Livingston detectors, and the detectors' averaged noise spectra in the same subband. We then compute the two $95\%$ confidence-level upper limits using the equation, and take the worse (i.e. the higher) among the two. 

This procedure has been validated by injecting simulated signals into LIGO and Virgo O2 and O3 data \cite{dicethesis}. In particular, this procedure has been shown to produce conservative upper limits with respect to those obtained with a much more computationally demanding injection campaign. It has been also verified that the upper limits obtained with a classical procedure based on injections are always above those based on the same analytical formula but using the minimum CR in each 1-Hz subband. The two curves, based, respectively, on the highest and the smallest CR, define a belt containing both a more stringent upper limit estimation and the sensitivity estimation of our search. When we will discuss the astrophysical implications of the search, we will always refer to the upper limits obtained using Eqn.~(67) in \cite{Astone:2014esa}, which are conservative with respect to limits derived from injections. This same procedure is being used for the O3 standard all-sky search for continuous waves from spinning neutron stars \cite{LIGOScientific:2022pjk}. {Eq. (67) of \cite{Astone:2014esa} gives the minimum detectable amplitude at a given confidence level, and has been derived by marginalizing over sky position and polarization of the source.}

\begin{figure*}[htb]
    \centering
    \includegraphics[width=1.5\columnwidth]{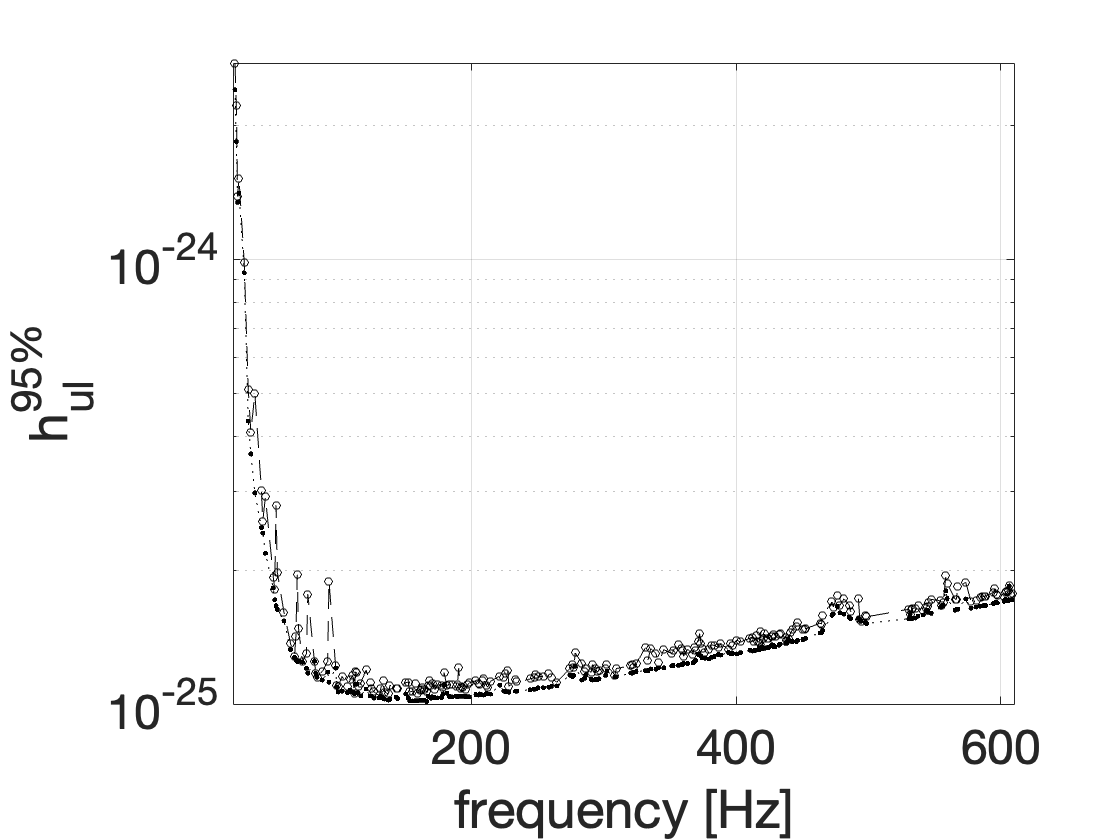}
    \caption{$95\%$ C.L. upper limits on the signal strain amplitude obtained from this search (circles, connected by the dashed line). Values are given in each 1-Hz subband. No value is given for the subbands where no outlier survives. The dots connected by a dashed line define the lower bound obtained using the minimum CR in each subband.}
    \label{fig:upper}
\end{figure*}

The resulting upper limits are shown in Fig.~\ref{fig:upper}, as circles connected by a dashed line. 
The minimum value is $1.04\times 10^{-25}$ at 130.5~Hz. In the same figure, the dots connected by the dashed line correspond to the lower bound computed using the minimum CR.
A comparison with previous all-sky searches, see e.g. Fig.~4 of \cite{CW_O3early} {or Fig. 6 of \cite{Steltner:2020hfd}}, shows that our conservative results are better than most of the past searches, including the recent early O3 analysis \cite{CW_O3early}. In particular, our minimum upper limit value improves upon that in \cite{CW_O3early} by $\sim 30\%$. The main factors that contribute to the improvement {relative to \cite{CW_O3early}} are: 1) the use of the full O3 C01 gated data, 2) the use of longer FFTs at least in a portion of the frequency band, and 3) the restricted spin-up/down range that could impact the maximum CR and thus the upper limit in each subband. On the other hand, the all-sky search described in \cite{CW_O3early}, as well as most of the other past searches, cover a significantly larger parameter space in both the spin-down/up and frequency ranges, and thus are sensitive to signals that are not considered in this analysis.
An O2 all-sky search reported in \cite{dergachev2021falconO2} produces upper limits slightly better than ours, by about $5-10\%$ on average (due to the use of a much longer coherent integration time of days), but obtained over a smaller parameter space and more limited in scope, being focused on low-ellipticity sources. See Sec.~\ref{sec:conclusions} for a more detailed discussion.  

We want to stress two important points here. First, our conservative procedure to compute upper limits, based on the maximum CR in each subband, produces strain values that are typically larger than the minimum detectable amplitudes at the same frequency. That is, the search sensitivity is expected to be better than the upper limits. 
Second, this is the first all-sky search that is optimized for frequency wandering signals, both at the level of outlier selection and follow-up. This allows us to achieve a better sensitivity to such signals compared to that achievable using the maximum possible FFT duration (which is the best choice for nearly monochromatic signals).  

\subsection{Astrophysical implications}
\label{sec:astro}
\begin{figure*}[htb]
    \centering
    \includegraphics[width=\columnwidth]{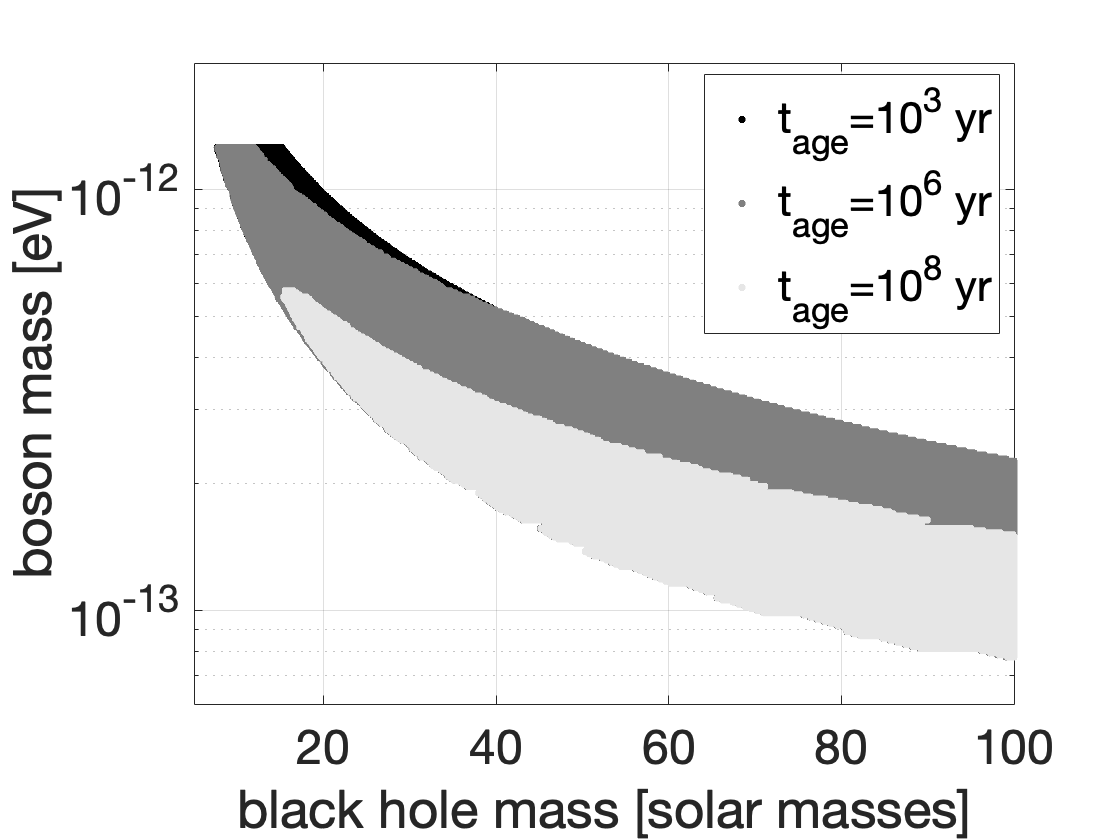}
    \includegraphics[width=\columnwidth]{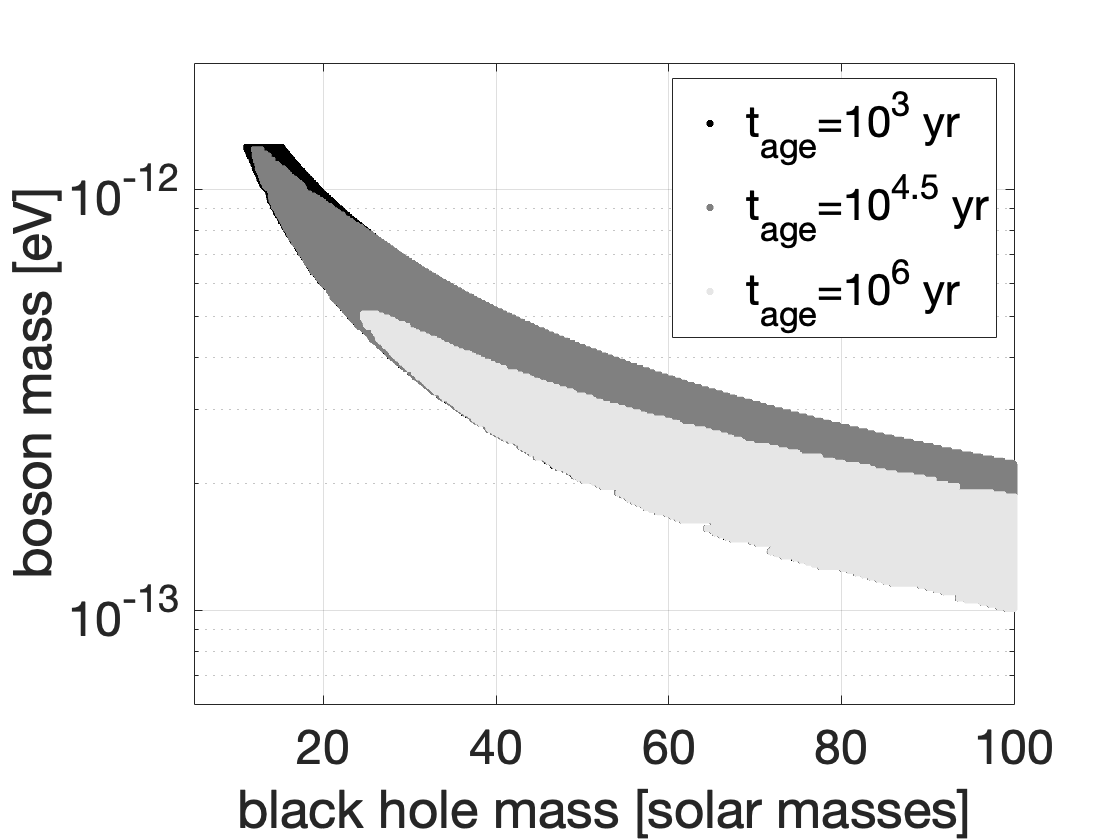}
    \caption{Exclusion regions in the boson mass ($m_b$) and black hole mass ($M_{\rm BH}$) plane for an assumed distance of $D=1$~kpc (left) and $D = 15$~kpc (right), and an initial black hole dimensionless spin $\chi_i=0.9$. For $D=1$~kpc, three possible values of the black hole age, $t_\mathrm{age}=10^3,~10^6,~10^8$ years, are considered; for $D=15$~kpc, $t_\mathrm{age}=10^3,~10^{4.5},~10^6$ years are considered.}
    \label{fig:exclude_chi09}
\end{figure*}

\begin{figure*}[htb]
    \centering
    \includegraphics[width=\columnwidth]{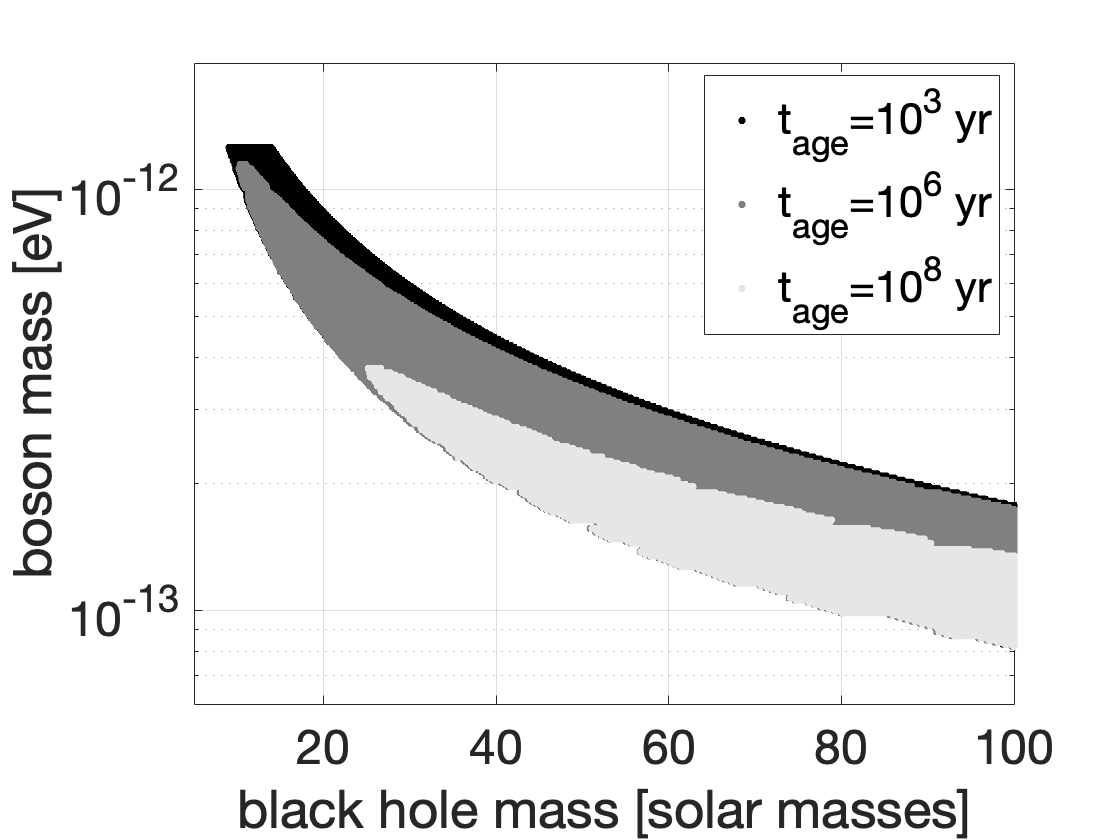}
    \includegraphics[width=\columnwidth]{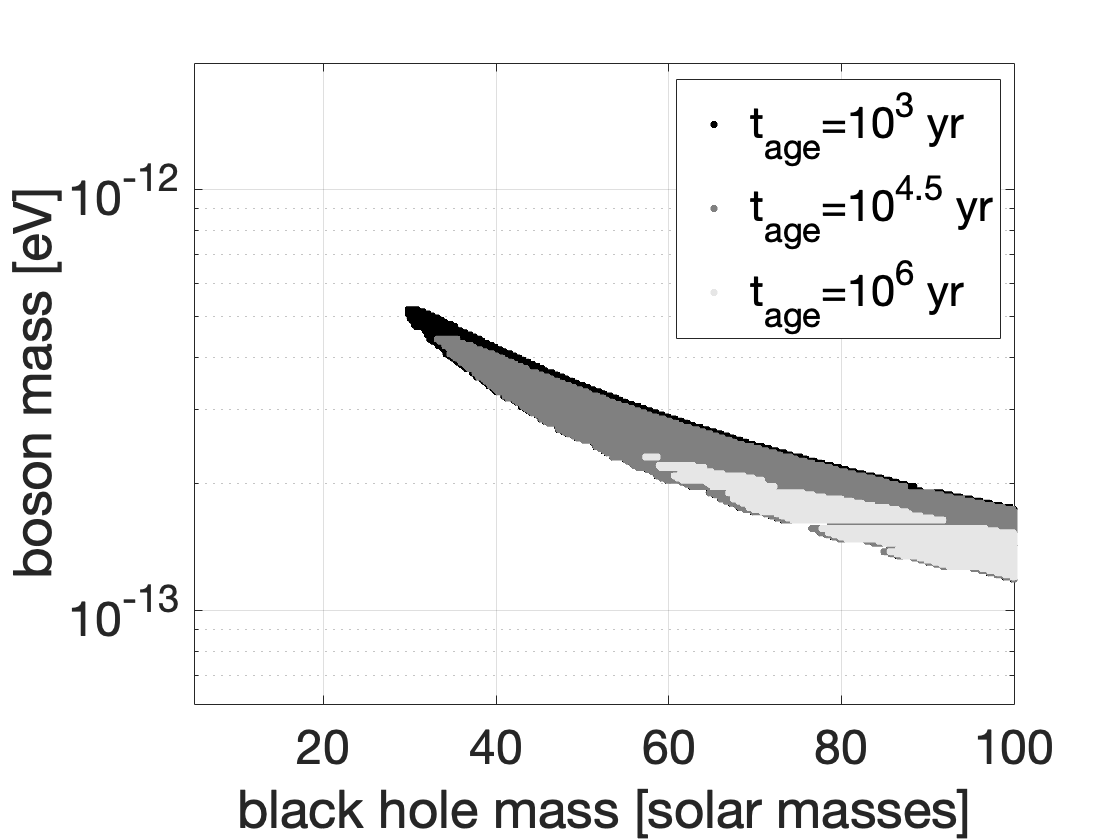}
    \caption{Same as Fig. \ref{fig:exclude_chi09} but for black hole initial spin $\chi_i=0.5$. The assumed distance is $D=1$~kpc (left), and $D=15$~kpc (right).}
    \label{fig:exclude_chi04}
\end{figure*}

The upper limits presented above can be translated into 
physical constraints on the source properties. We interpret the results in two different ways. First, similar to what has been done in \cite{palomba2019direct}, we compute the exclusion regions in the boson mass and black hole mass plane, assuming fixed values of the other relevant parameters, namely the distance to the system $D$, the initial dimensionless spin of the black hole $\chi_i$, and a time equal to the age of the cloud $t=t_\mathrm{age}$. Indeed, from Eqs.~(\ref{eq:h0}) and (\ref{eq:hoft}), once $D$, $\chi_\mathrm{i}$, and $t_\mathrm{age}$ are fixed, the signal strain amplitude depends only on the boson mass $m_b$ and black hole mass $M_\mathrm{BH}$. Thus for each pair of $(m_b,~ M_\mathrm{BH})$, we can exclude the presence of a source with those assumed parameters, if it would have produced a signal whose strain is larger than the upper limit at a given frequency in figure \ref{fig:upper}. 
In our search, the $\alpha$ values probed roughly fall in the range of [0.02, 0.13]. Eqs.~(\ref{eq:taugw}) and (\ref{eq:h0}) do not hold in the range of $\alpha \gtrsim 0.1$, as shown in Fig. 2 of \cite{Brito:2017zvb}. Specifically, we look at the Brito+ curve derived analytically compared to the black curve obtained from numerical simulations, and see that they differ in energy by a factor of 3 at the largest $\alpha$ considered in this search, $\alpha~0.15$.
This implies that we underestimate $\tau_\mathrm{gw}$ up to a factor of 3 and overestimate $h_0$ up to a factor of $\sqrt{3}$ (at $\alpha\simeq 0.13$). 
Thus we correct Eqs.~(\ref{eq:taugw}) and (\ref{eq:h0}) by these factors such that the resulting exclusion regions are slightly conservative in the whole parameter space.
In Fig.~\ref{fig:exclude_chi09}, the exclusion regions are plotted for $D = 1$~kpc (left) and $D = 15$~kpc (right), assuming a high initial spin of $\chi_i=0.9$. For each distance assumption, three different values of the black hole age are considered. 
In Fig.~\ref{fig:exclude_chi04}, the exclusion regions are shown for the same parameters as before, except that a lower initial spin value, $\chi_i=0.5$ is assumed. 
In both cases, as expected, the constrained region is smaller when the source is assumed to be at a farther distance as, in this case, the signal amplitude at the detector is smaller. 
For any given black hole mass, these results improve upon the constraints obtained in Advanced LIGO O2 data \cite{palomba2019direct} for lower boson masses, while they are slightly less constraining for higher boson masses. This is because for a given black hole mass, higher boson masses correspond to higher signal spin-up (see Eq. (\ref{eq:fanni})), and we simply do not cover a large spin-up range. 
On the other hand, the constraints described in \cite{palomba2019direct} were obtained from the results of a search not specifically designed for boson clouds. In particular, restricting the exploration to the parameter space relevant for the expected boson cloud signals significantly reduces the trial factor with a consequent implicit gain in the sensitivity with respect to standard wider parameter space searches.

\begin{figure*}[htb]
    \centering
    \includegraphics[width=1\columnwidth]{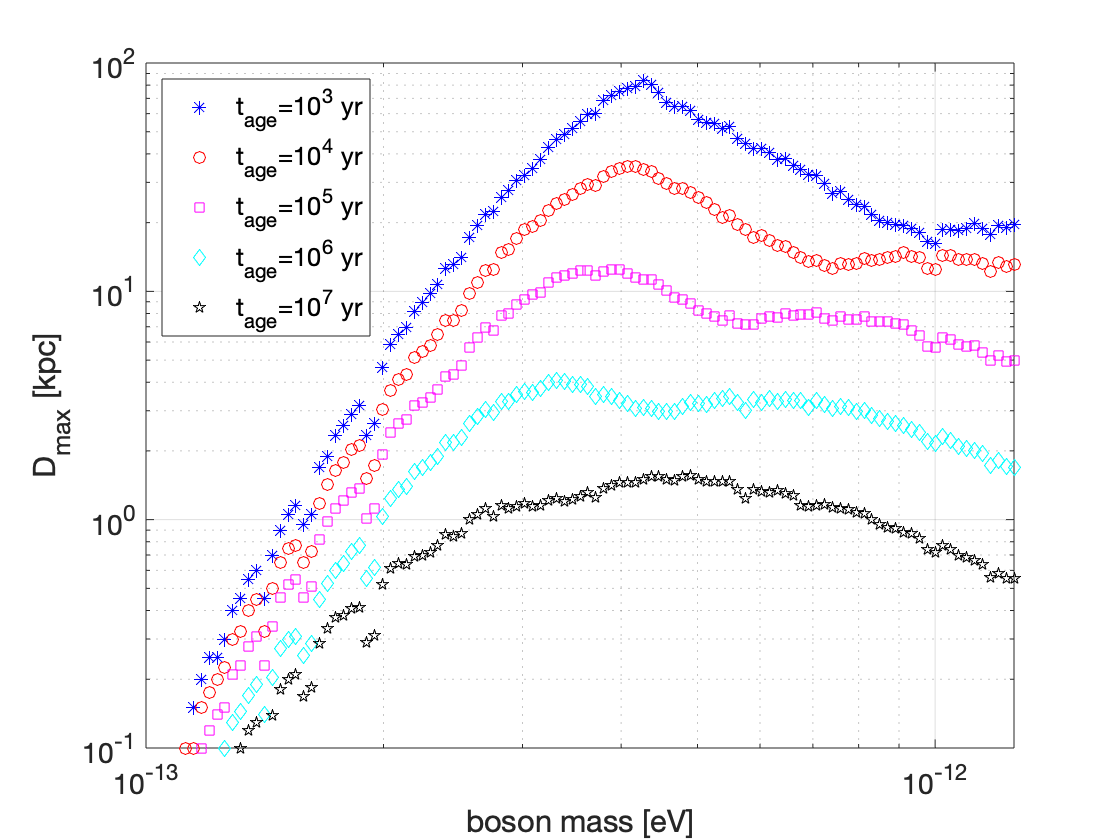}
    \includegraphics[width=1\columnwidth]{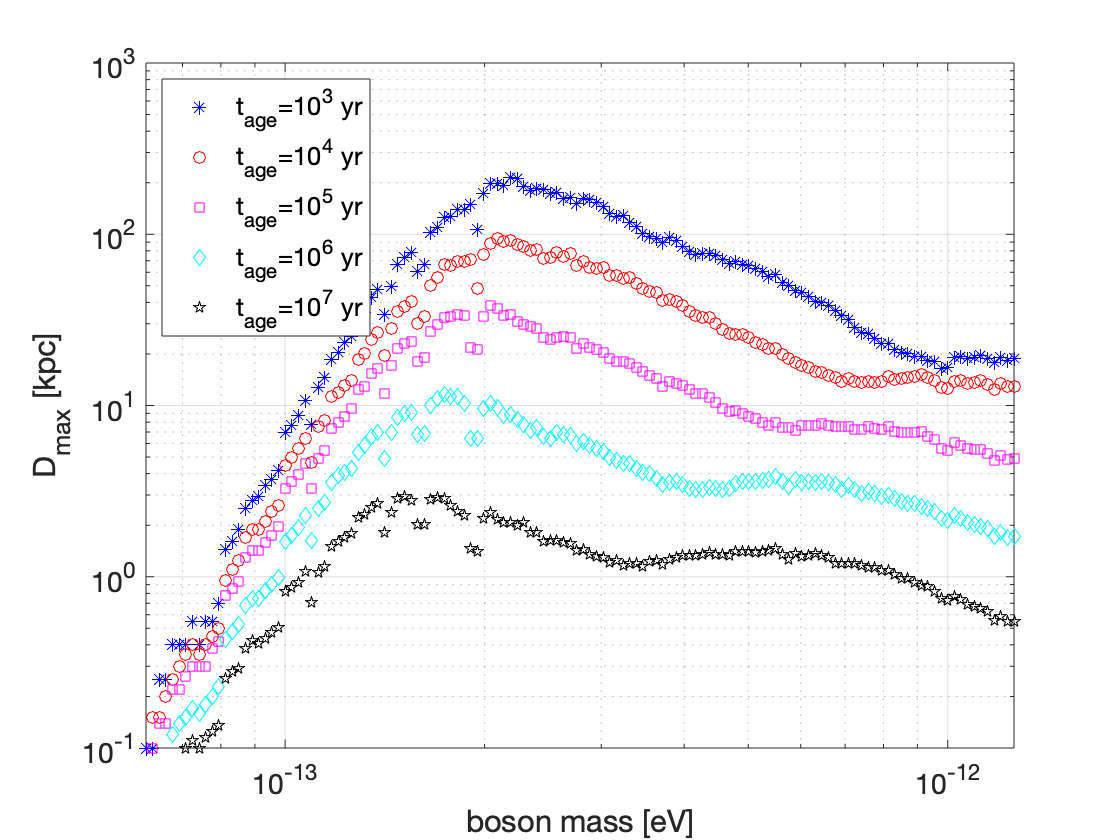}
    \caption{Maximum distance at which at least $5\%$ of a simulated population of black holes with a boson cloud would produce a gravitational-wave signal with strain amplitude larger than the upper limit in the detectors. The left plot refers to a maximum black hole mass of 50$M_{\odot}$, while the right plot to a maximum mass of 100$M_{\odot}$. The different colored markers correspond to different system ages, ranging from $10^3$ years to $10^7$ years, as indicated in the legend. The alignment of points for different ages at the smallest boson masses (and distances) is the result of a discretization effect due to the finite size grid used in distance.}
    \label{fig:maxdist}
\end{figure*}

A second type of interpretation, as shown in Fig.~\ref{fig:maxdist}, is represented by the maximum distance $D_\mathrm{max}$ at which we can exclude, as a function of the boson mass, the presence of an emitting system of a given age $t_\mathrm{age}$, assuming a population of black holes. The black hole population has been simulated using a Kroupa mass distribution,  with probability density described by $f(m)\propto m^{-2.3}$ \cite{elbert2017}, with two different ranges, of $[5,~50]M_\odot$ and $[5,~100]M_\odot$, and a uniform initial spin distribution in the range of $[0.2,~0.9]$. The criterion to compute $D_\mathrm{max}$ at a given boson mass is that at least $5\%$ of the simulated signals would produce a strain amplitude larger than the upper limit at the corresponding frequency in the detectors, and thus would have been detected by our search. This choice of 5\% should be conservative, given the large black hole population in the galaxy (10$^7$-10$^8$) from which we seek a single signal. As before, we correct Eqs.~(\ref{eq:taugw}) and (\ref{eq:h0}) by a factor of 3 and $\sqrt{3}$, respectively, to obtain more accurate $\tau_{\rm gw}$ and $h_0$ estimates at $\alpha \gtrsim 0.1$ and thus slightly conservative constraints in the full range. As expected, on average when the maximum black hole mass is smaller, the maximum distance is also smaller, as the signal strain increases with a high power of the black hole mass.
Instead of focusing on the specific properties of the emitting system, as for the exclusion regions previously discussed, this constraint on distance depends on the chosen Kroupa mass and uniform spin distributions of the black holes, and thus it reflects the ensemble properties of the assumed black hole population. 

Overall, the distance constraints allow us to draw semi-quantitative conclusions on the possible presence of emitting boson clouds in our Galaxy. For instance, young systems, with $t_{\rm age}$ smaller than about $10^3$ years are disfavored in the whole Galaxy for boson masses above about $2.5\times 10^{-13}$~eV for a maximum black hole mass of 50$M_{\odot}$ and above about $1.2\times 10^{-13}$~eV for a maximum black hole mass of 100$M_{\odot}$. As expected, older systems, expected to be more abundant, are more likely to be ruled out only at smaller distances, as they produce on average weaker gravitational-wave signals. As an example, bosons with masses between $\sim 2\times 10^{-13}$--$8\times 10^{-13}$~eV ($\sim 10^{-13}$--$8\times 10^{-13}$~eV) would be excluded within a distance of 1~kpc from Earth for a maximum black hole mass of 50$M_{\odot}$ (100$M_{\odot}$). 

The general shape of the maximum distance curves is a result of the combination of different factors pushing in different directions: higher boson masses tend to produce signals with stronger initial amplitudes, but also faster decay rates as a function of $t_{\rm age}$. 
Towards the lower $m_b$ end in the plot, $D_{\max}$ increases with $m_b$ because the effect of the increasing initial signal amplitudes dominates; towards the higher $m_b$ end, $D_{\max}$ tends to decrease again since the faster signal decay rate dominates. 
At the same time, for higher boson masses, lower black hole masses are required to form a cloud producing signals that last long enough to be detectable. 
Hence the black hole mass distribution (larger population at lower $M_{\rm BH}$) comes into play as well.
In addition, the results are weighted by the frequency-dependent detector sensitivity (reflected in the upper limit curve in Fig.~\ref{fig:upper}), which also affects the final shape of the $D_{\max}$ curves.

For both of the constraints presented above, we do not take into account factors that are largely uncertain but could also be relevant to the interpretation, e.g., the black hole spatial distribution or the formation rate. We defer the integration of these factors to future work when they are better constrained. 

Concerning the interpretation of our results with respect to the assumed boson self-interaction, the search covers a small range of spin-up values (see Sec~\ref{sec:method} and, in particular, Fig.~\ref{fig:spindownstep}), mainly corresponding to a regime of small boson self-interaction, i.e. one in which the spin-up is dominated by Eq~(\ref{eq:fanni}). For a subset of smaller boson and black hole masses, the spin-up dominated by Eq~(\ref{eq:fdot_tr}) in the intermediate regime is also covered in the search. However, as we have clarified in Sec~\ref{sec:model}, the signal duration is expected to be shorter in this intermediate regime, and the signal amplitude is smaller due to the smaller boson and black hole masses [see Eq~(\ref{eq:h0})]. Thus the signals in the intermediate regime are less likely to be detected compared to those in the small self-interaction regime. 

\section{Conclusions}\label{sec:conclusions}
This paper describes the first all-sky search tailored to the predicted continuous-wave emission from scalar boson clouds around spinning black holes. 
We cover a frequency range between 20~Hz and 610~Hz, and a small frequency-dependent spin-up range corresponding to the small self-interaction regime. 

We use a multiple frequency-resolution approach, in order to optimize the sensitivity for signals characterized by a slightly wandering frequency, for which a search based on the fixed maximum possible Fourier transform duration would cause a sensitivity loss. No candidate survives the multi-step follow-up procedures we have implemented.

Following up outliers in this analysis was significantly more challenging due, in part, to not having a spin-down or spin-up range to consider, typical of standard all-sky \cwh analyses, for which instrumental artifacts lead to clusters over extended ranges in source spin-down. Establishing an instrumental source of such clusters is more straightforward in standard analyses than in this analysis. Manual checks by visual inspection of many spectra had to be performed, along with additional methods necessary to reject an outlier due to a transient spectral artifact in H1 (see section \ref{sec:fuw1}). Therefore, this work not only motivates further searches for boson cloud systems, but also refined methods to mitigate or conclusively veto persistent or transient spectral disturbances at the level of detector commissioning and characterization.

The resulting upper limits are significantly better than those obtained in previous all-sky searches for continuous waves, including a recent search in the early O3 data \cite{CW_O3early}, although our search typically covers a much smaller spin-down/up range with respect to those.
On the other hand, our upper limits are slightly worse than those obtained in the O2 search carried out with the Falcon pipeline \cite{dergachev2021falconO2}, which however was mainly focused on low-ellipticity sources, and thus covered 
a much smaller spin-down/up range than we do (from a factor of $\sim 4$ at lower frequencies up to a factor of $\sim 18$ at higher frequencies), and moreover did not account for the possibility of signal frequency wandering. The same pipeline was also run from 500-1700 Hz \cite{Dergachev:2020fli} with the same caveats, and although the limits presented in both of these searches are quite stringent, they rely on low-ellipticity sources emitting almost monochromatic \gws. {Furthermore, the results in the most recent run of the Falcon pipeline in the 500-600 Hz region using O3a data surpass ours, and could also be interpreted in terms of \gwh emission from boson clouds \cite{Dergachev:2022lnt}.}

Code improvements are possible to make the search more sensitive and computationally more efficient. In this way, future boson cloud searches using the basic methodology adopted in this paper will expand the parameter space, covering a wider frequency band and a larger spin-up range, as well as search for vector boson cloud signals that could have much higher spin-ups than those considered here. This, together with the expected detector improvements in upcoming runs of Advanced LIGO, Advanced Virgo and KAGRA detectors,  will significantly increase the chance of detecting gravitational radiation from these interesting sources, or at least to better constrain the parameter space of these systems.

\appendix

\section{Hardware injections}
\label{sec:hi}

\begin{figure}[htb]
    \centering
    \includegraphics[width=\columnwidth]{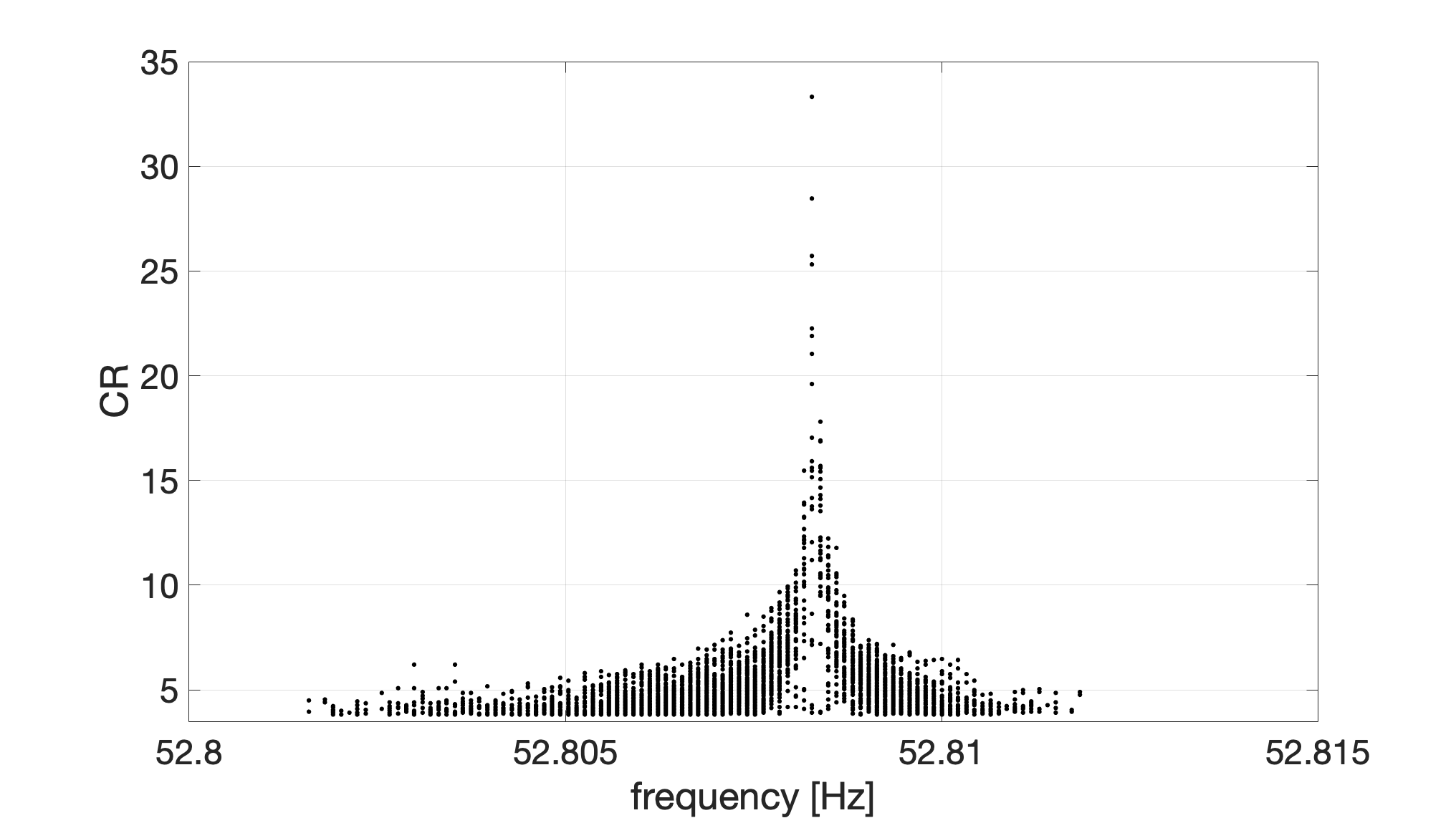}
    \caption{Critical ratio of the candidates produced by {\it hardware injection} P5 in H1 data and found in the search with $W=1$. The distance of the highest CR candidate from the injected signal is about 0.7.}
    \label{fig:cand_P5}
\end{figure}

Both H1 and L1 data contain 14 simulated continuous-wave signals, called {\it hardware injections}, added via hardware for testing purposes. Only one of them, named P5, is within the searched frequency and $\dot{f}$ range and has an amplitude high enough to be detectable in this search. Indeed we find it with a maximum $\rm{CR}\simeq 33$ and 36, respectively in the two detectors (and with an averaged dimensionless distance $d \approx 0.7$). As this injected signal is very strong, it produces tens of thousands of outliers at slightly different points in the parameter space.
Thus we remove these outliers caused by the hardware injection P5.
Fig. \ref{fig:cand_P5} shows the outliers found in H1 data caused by this injected signal.

\section{Follow-up methodology}
\label{sec:fumethods}

Here we provide details about the two methods used for outlier follow-up. The first one, applied to outliers with $W=1$, is more suitable for monochromatic or nearly monochromatic signals, while the second works well for signals characterized by some amount of frequency fluctuations.
In principle, both methods can be applied to all outliers. 
The sensitivities of the two methods cannot be directly compared since the two methods assume different signal models. To keep the number of outliers at a practical level for both methods, we only use one method for each category. 
We assume that for outliers with $W=1$, the impact from any potential random frequency fluctuation is negligible. 

\subsection{Follow-up for outliers with $W=1$}
\label{sec:fu_FH}
\begin{figure}[htb]
    \centering
    \includegraphics[width=\columnwidth]{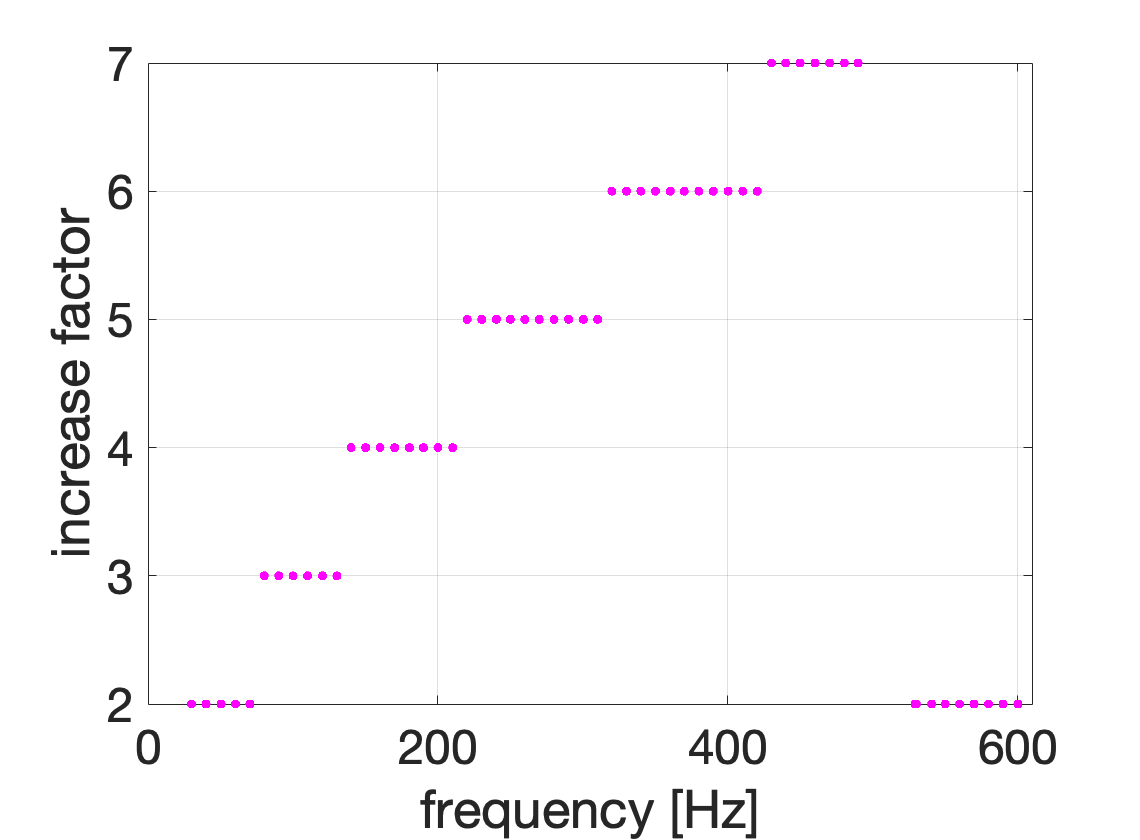}
    \caption{FFT length increase factor, with respect to the main search FFT length, as a function of frequency for the first follow-up stage (FU1). This factor is reduced at the highest frequencies in order to keep the computer memory used by the analysis jobs under control.}
    \label{fig:fftgain}
\end{figure}

The follow-up for monochromatic or nearly monochromatic signals consists of several steps, applied to each outlier, and is described in detail in \cite{pisarski2019all}. First, given the frequency and sky coordinates of an outlier, the Doppler modulation of a possible \gwh signal is removed from the data by heterodyne\footnote{This is a standard technique that involves multiplying the data by a phase factor $\exp[{-j\phi_d(t)}]$, where $\phi_d(t)$ is the time-dependent phase evolution induced by the Doppler effect.}.
This allows us to build a new set of FFTs, with a longer duration with respect to the original value, as well as the corresponding peakmap. The factor by which the FFTs duration is increased is a function of the frequency and is shown in Fig.~\ref{fig:fftgain}. 
It is chosen as a compromise between the need to improve sensitivity and to keep the computational cost of the follow-up under control. Using different lengthening factors at different frequencies aims to produce similar durations for the new set of FFTs, which would correspond to similar follow-up sensitivity, taking into account that the initial FFTs are shorter at higher frequencies (see Fig. \ref{fig:tfft}). In practice, due to computer memory limitations, we have to limit the increase at the highest frequencies between 500~Hz and 600~Hz. 

A follow-up volume, defined by $\pm 3$ frequency bins and $\pm 3$ bins for each sky coordinates (these bins, called {\it coarse} bins, are those of the initial search setup) around the outlier, is considered. 
For each sky position and frequency in the follow-up volume, by linear extrapolation of the detector velocity vector, we evaluate the residual Doppler modulation (with respect to the center of the grid), which is corrected in the peakmap by properly shifting each peak. This extrapolation procedure is orders of magnitude faster than the exact Doppler correction and is accurate as long as the follow-up volume is sufficiently small (see \cite{pisarski2019all}). 

The Frequency-Hough algorithm \cite{Astone:2014esa} is then applied to the resulting ensemble of corrected peakmaps (covering $\pm 1$ coarse bin in $\dot{f}$). It maps the detector time-frequency plane of the peakmap to the source frequency-$\dot{f}$ plane.  
The absolute maximum identified among all frequency-$\dot{f}$ Frequency-Hough histograms provides the refined parameters of the original outlier. This procedure is applied to each element of coincident pairs of outliers. 

Next, we apply a series of vetoes to each pair of refined outliers, detailed as follows.
First, for each detector, a new peakmap is computed using the data coherently corrected with the refined parameters of the outlier and then projected onto the frequency axis. We take the maximum of this projection in a range of $\pm 2$ coarse bins around the outlier frequency\footnote{The $ \pm 2 $ bins used at this level do not have any particular relation with the $\pm 3$ bins used to define the follow-up volume. The latter choice is due to the uncertainty on the parameters of the outlier, while the former is chosen to estimate the outlier significance.}.

The rest of a 0.2-Hz band around the outlier, which we consider as the “off-source” region, is divided into a frequency-dependent number of intervals of the same width (in practice, this number varies from about 600 at the lowest frequencies, to about 100 at the highest frequencies). We take the maximum of the peakmap projection in each of these intervals and sort all these maxima in a decreasing order. 
We keep the outlier if it ranks in the $1\%$ top list in both detectors. 

\begin{figure}[htb]
    \centering
    \includegraphics[width=\columnwidth]{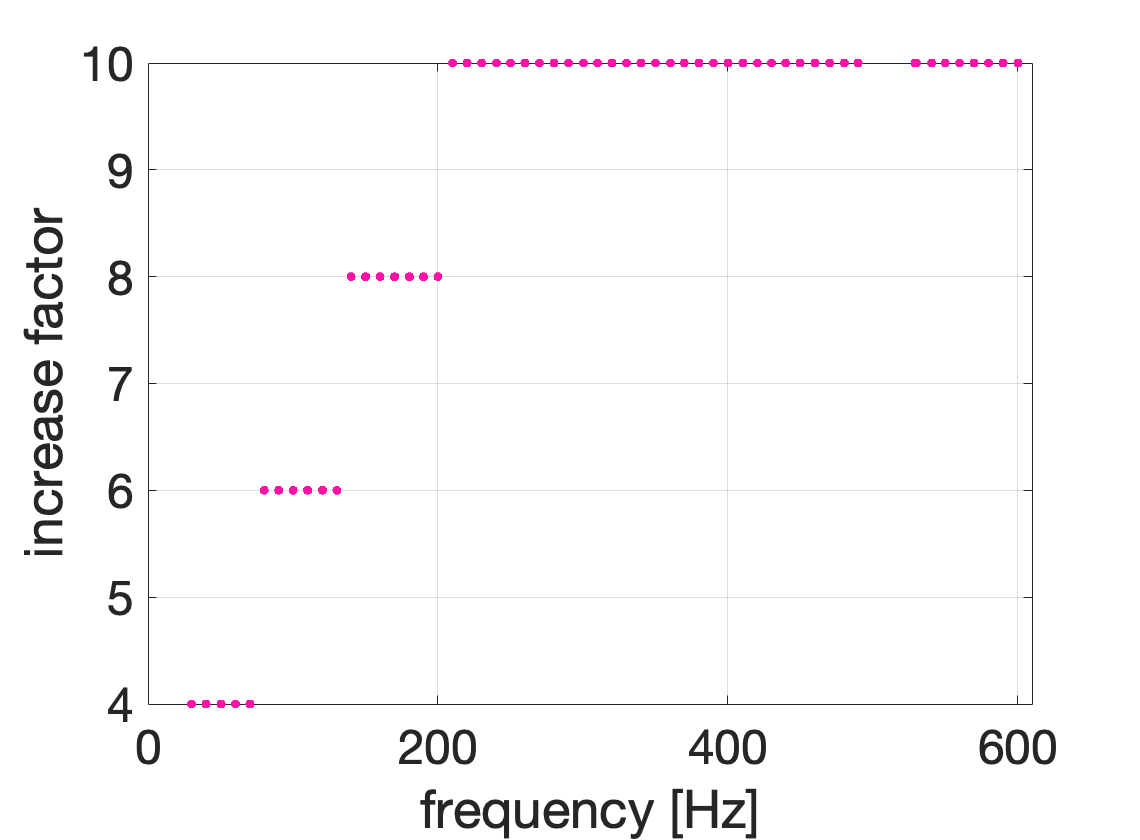}
    \caption{FFT length increase factor, with respect to the main search length, as a function of frequency for the second follow-up stage (FU2).}
    \label{fig:fftgain2}
\end{figure}

Surviving outliers are then subject to three further tests. 
The first is a test based on the consistency of the outlier CR in the two detectors, similar to that already used in the main search. The second is a test in which an outlier is discarded, if its CR after the follow-up (in both detectors) drops compared to the original CR from the main search. Finally, we discard coincident outliers with $d>6$ after the follow-up, indicating that they can no longer be considered as coincident. 
Outliers surviving this first follow-up round (FU1) are subject to a second round (FU2) in which the same procedure is used but the FFT duration is further increased. Fig.~\ref{fig:fftgain2} shows the lengthening factor in FU2, as before with respect to the initial duration.
The follow-up volume around each outlier consists, as before, of $\pm 3$ coarse frequency bins, $\pm 3$ bin for each sky coordinate and $\pm 1$ $\dot{f}$ bin around the outlier (note that these are {\it coarse} bins of the initial main search).
If needed, a third follow-up stage (FU3) along the same line, is applied. The follow-up volume is the same as in previous steps, while the FFT length increase is chosen specifically depending on the outlier parameters.
Any outliers surviving these follow-up steps will be eventually analyzed with ad-hoc methods.

\subsection{Follow-up for outliers with $W>1$}
\label{sec:fu_viterbi}

Outliers found with $W>1$ are expected to have some level of frequency evolution that leads to power leakage into adjacent frequency bins.
We use the Viterbi method, based on a HMM tracking scheme, which has been used in various continuous-wave searches~\cite{isi2019directed,Suvorova:2016rdc,Sun:2017zge,Sun:2018hmm,Sun:2019mqb}.
The Viterbi method can efficiently track secular evolution as well as random fluctuation in the signal frequency, and thus is an ideal follow-up tool to verify candidates with $W>1$. 

For each outlier identified with $W>1$, we first estimate the maximum $\dot{f}$ value assuming that all the signal power is distributed in $W$ frequency bins,
\begin{equation}
    \dot{f}_{\rm max} = \frac{W}{2T_{\rm obs}T_{\rm FFT}},
\end{equation}
and select a coherent time $T_{\rm coh}$ such that the $\dot{f}_{\rm max}$ can be covered in the Viterbi follow-up search. See details about how to select $T_{\rm coh}$ in Section IV A of \cite{isi2019directed}. In this follow-up study, three different values of $T_{\rm coh} = 1$d, 2d, and 4d, are used, with shorter $T_{\rm coh}$ applied to larger $\dot{f}_{\rm max}$. For $T_{\rm coh} = 1$d, 2d, and 4d, a narrow frequency band of 0.5~Hz, 0.125~Hz, and 0.03125~Hz, centered at the outlier frequency, is searched, respectively. The best estimated sky position in the main search is used for the follow-up analysis.
This follow-up procedure consists of three steps. 

First, we search in the combined data from LIGO Hanford and Livingston using the Viterbi method with the $T_{\rm coh}$ and frequency band configurations listed above. The detection score $S$ is defined, such that the log likelihood of the optimal Viterbi path equals the mean log likelihood of all paths ending in different frequency bins plus $S$ standard deviations in each band searched~\cite{Sun:2017zge}. We discard outliers which are not found in the Viterbi follow-up, either $S$ is below the threshold (corresponding to 1\% false alarm probability) or the optimal Viterbi path is completely irrelevant to the original outlier (returned at a nonoverlapping frequency). 
Second, we eliminate the outliers whose significance is higher when analyzing one detector only than combining the two, while searching the other detector alone yields $S$ below the threshold, i.e., the outlier is dominantly louder in a single detector. 
Third, we carry out manual inspections for any remaining outliers that survive the first two steps.

\section{Candidates and follow-up results}
\label{app:cand}

\begin{figure*}[htb]
    \centering
    \includegraphics[width=\columnwidth]{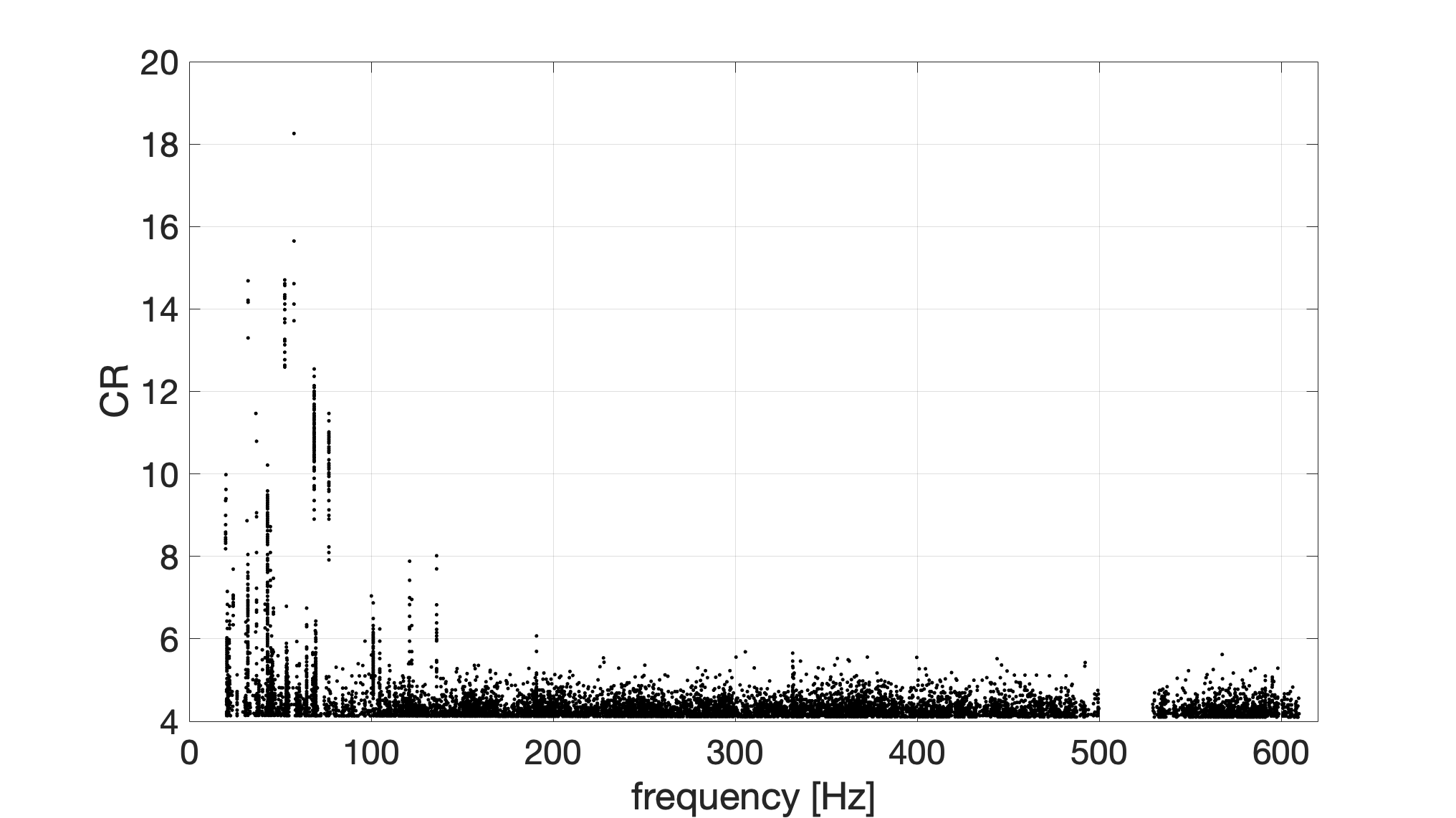}
    \includegraphics[width=\columnwidth]{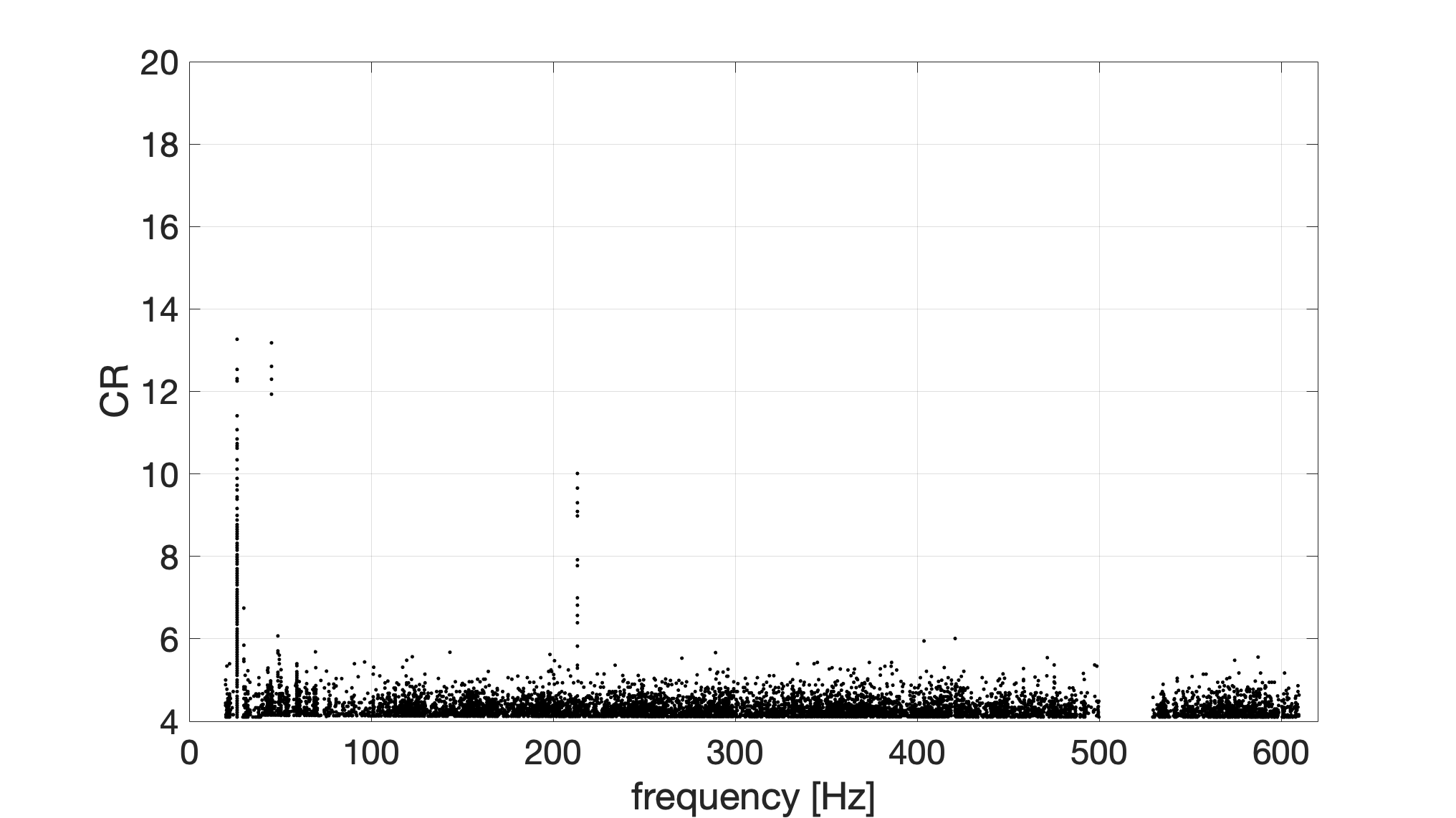}
    \caption{Critical ratio of the main search H1 (left) and L1 (right) outliers with $W=1$.}
    \label{fig:cand_W_1}
\end{figure*}

\begin{figure*}[htb]
    \centering
    \includegraphics[width=\columnwidth]{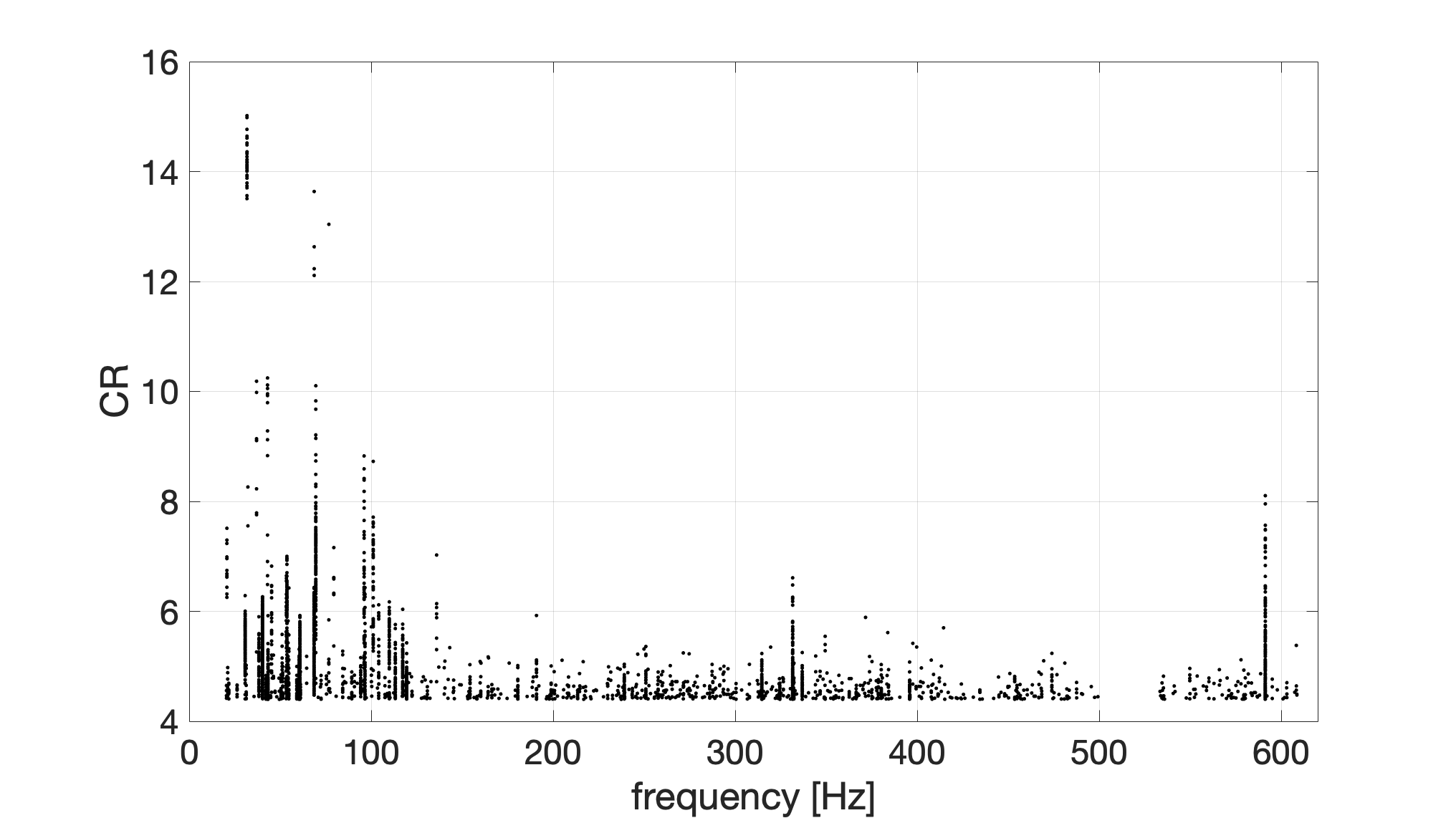}
    \includegraphics[width=\columnwidth]{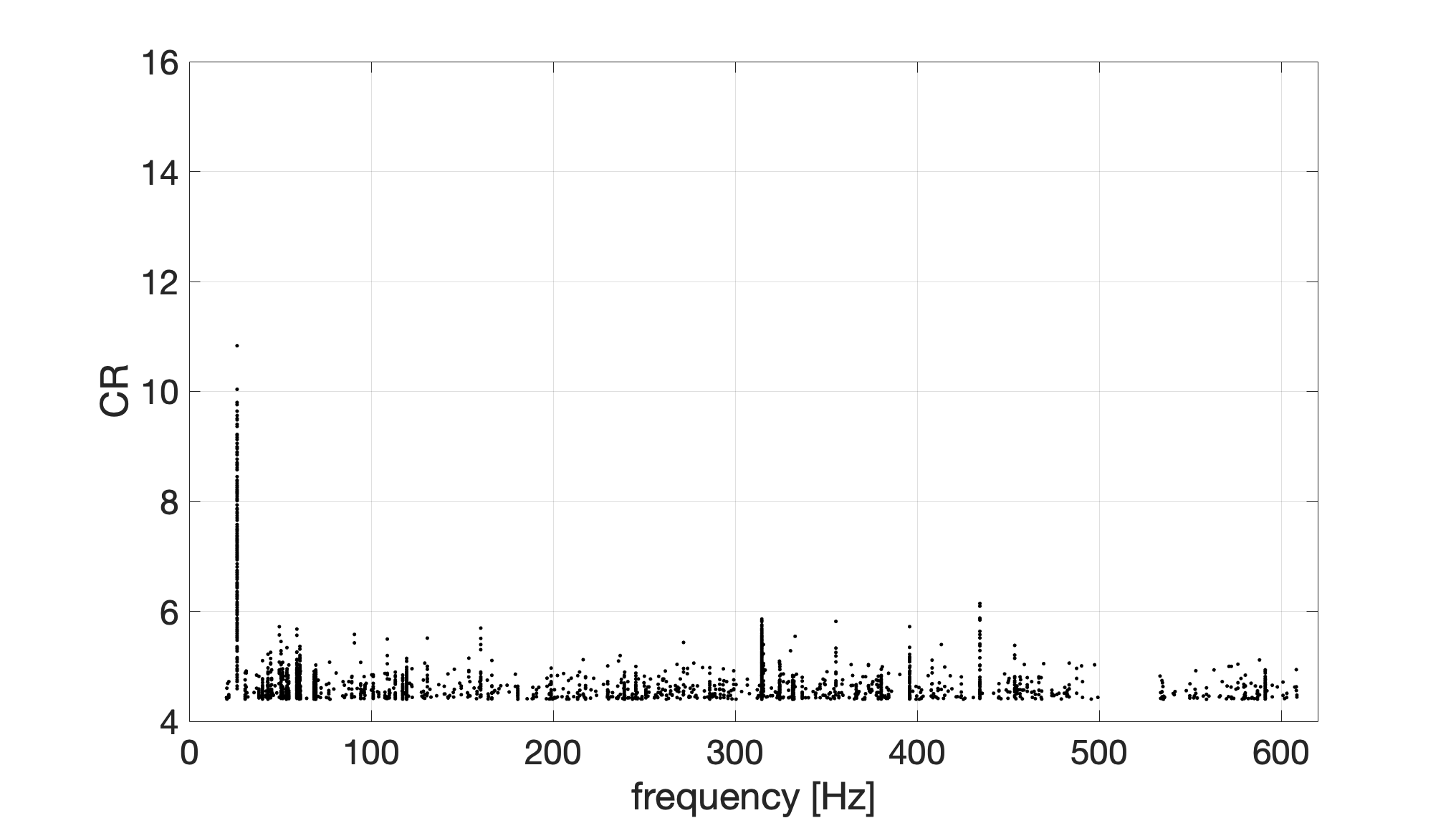}
    \caption{Critical ratio of the main search H1 (left) and L1 (right) outliers with $W>1$.}
    \label{fig:cand_W_2}
\end{figure*}

The outliers produced by the main search, with average window $W=1$ and $W>1$, are shown, respectively, in Figs.~\ref{fig:cand_W_1} and \ref{fig:cand_W_2} in terms of their CR as a function of the frequency.
It is clear that the majority of the strongest outliers is due to H1 data, and is concentrated in the lower frequency range below $\sim 150$~Hz. In the following, we present details of the follow-up results for the outliers.

\subsection{Follow-up results for outliers with $W = 1$} \label{sec:fuw1}

For outliers with $W=1$, we apply the follow-up procedure described in Appendix~\ref{sec:fu_FH}. 

We start with the first follow-up stage (FU1), which produces outliers with refined parameters, on which various vetoes are applied.
The application of the significance veto removes most of the outliers, with only 160 surviving. Among these, 14 are discarded as their CR decreases, and an additional 10 are eliminated because they yield a post-follow-up distance of $d>6$ (i.e., they are no longer coincident). 

\begin{figure}[htb]
    \centering
    \includegraphics[width=\columnwidth]{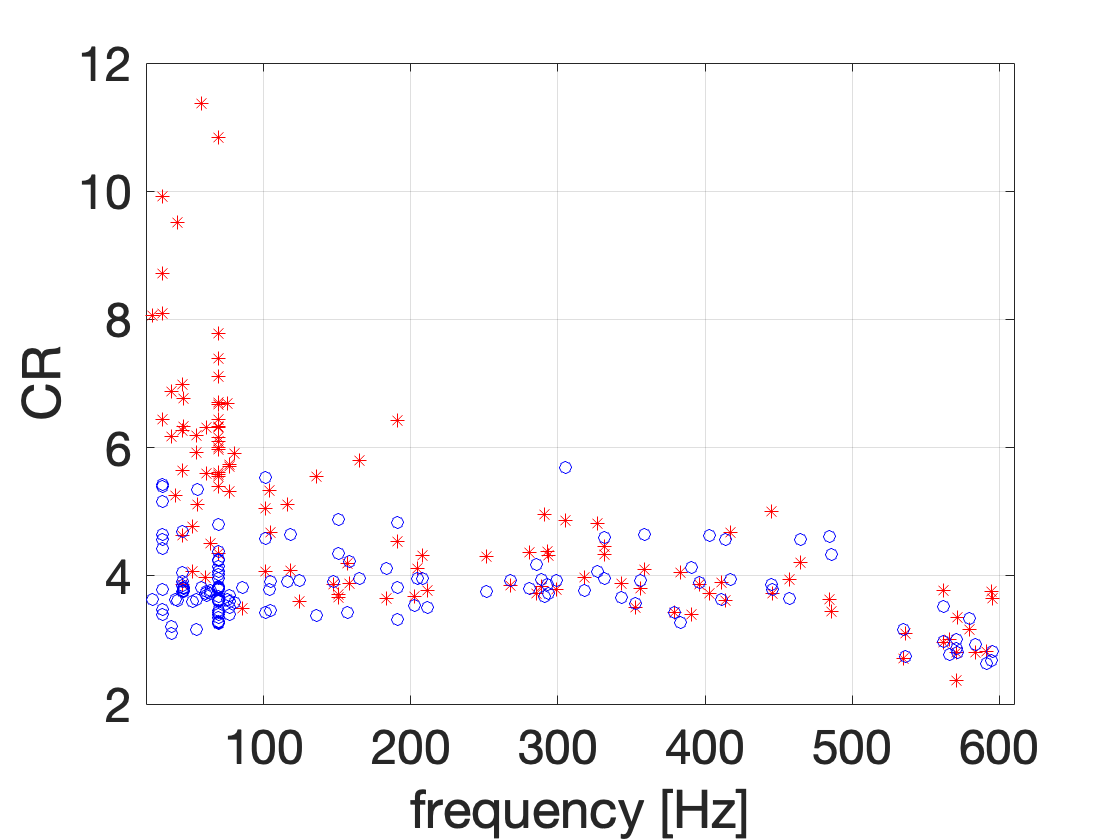}
    \caption{Critical ratios of outliers passing FU1 follow-up stage as a function of their frequencies (circles: Livingston, stars: Hanford).}
    \label{fig:FU1}
\end{figure}

\begin{figure}[htb]
    \centering
    \includegraphics[width=1.05\columnwidth]{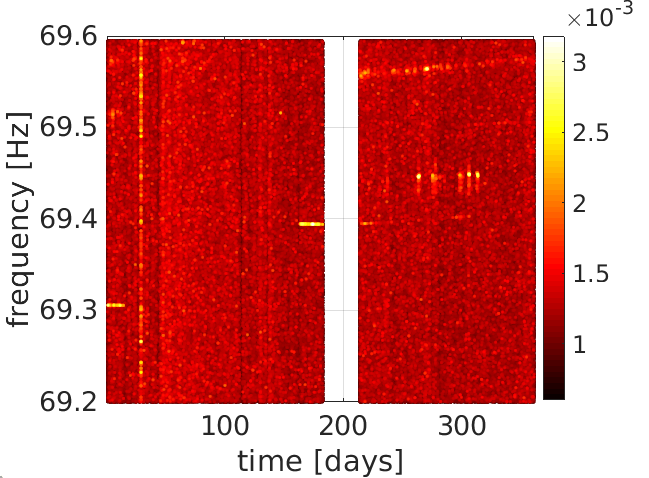}
    \caption{Time-frequency peakmap of H1 data after FU1 follow-up stage showing a distinct broad disturbance appearing in the second half of the run and responsible for the 22 outliers identified with frequency $\sim 69.55$~Hz. Time is measured in days since the stsrting of the run. The colorbar represents a power spectrum in units of $10^{-40}$ [1/Hz].}
    \label{fig:peakmap_69}
\end{figure}

The CRs of the 136 outliers passing FU1 are shown, as a function of frequency, in Fig.~\ref{fig:FU1}.
By visual inspection of the refined time-frequency peakmaps produced in FU1, we are able to conclude that many of the strongest outliers are not compatible with an astrophysical signal. In particular, this happens for one outlier at 24.20~Hz, six at $\sim 31.14$~Hz, two at $\sim 37.31$~Hz, one at $\sim 41.72$~Hz, one at $\sim 57.59$~Hz, three at $\sim 69.39$~Hz and 22 at $\sim 69.55$~Hz. 
As an example, we show in Fig.~\ref{fig:peakmap_69} the refined peakmap built from H1 data, indicating a clear transient disturbance in the second half of the run and responsible for the outliers at $\sim 69.55$~Hz.

The remaining 97 outliers are analyzed in the second follow-up stage (FU2), which reduces the number of surviving candidates to three (91 are removed by the significance veto and 3 by the distance veto).

\begin{figure}[htb]
    \centering
    \includegraphics[width=\columnwidth]{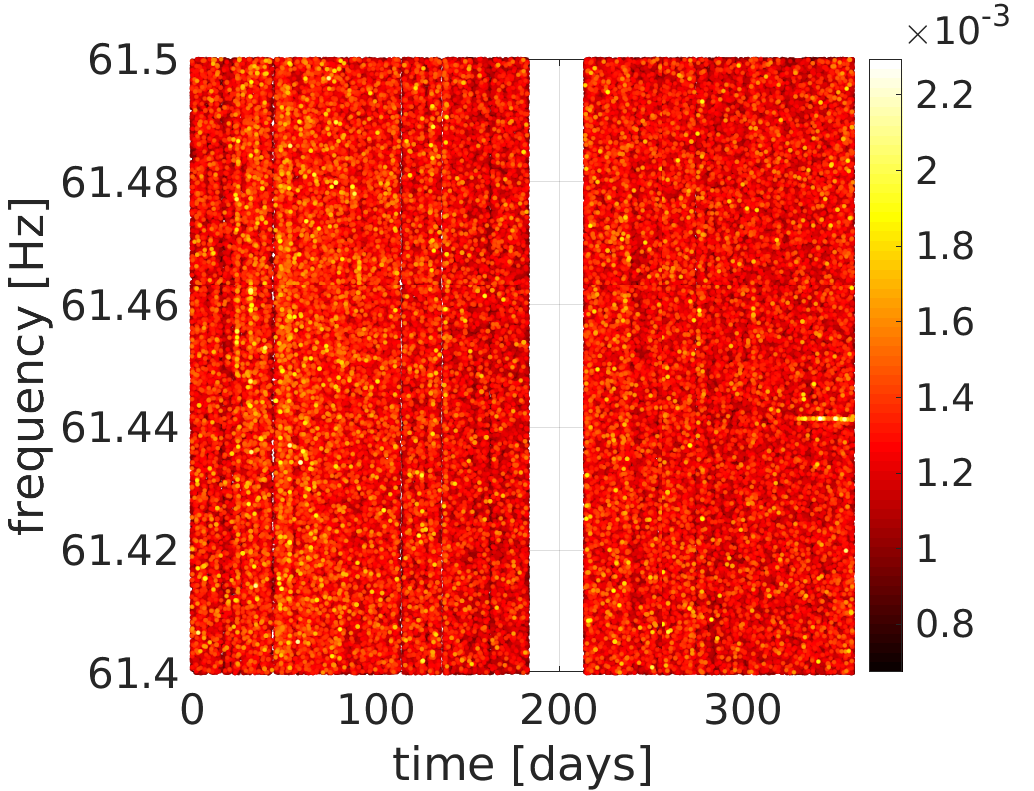}
    \caption{Time-frequency peakmap of H1 data after the second follow-up stage (FU2) around the outlier with frequency $\sim 61.44$~Hz. Time is measured in days since the starting of the run. A transient disturbance, starting from $\sim 20$ days before the end of the run, is clearly visible. The color bar represents power spectrum in units of $10^{-40}$ [1/Hz].}
    \label{fig:cand_61_H}
\end{figure}

The final three remaining outliers have average parameters (averaged values of the parameters obtained from the two detectors) shown in Table~\ref{tab:candidates}. 
\begin{table*}[t]
    \begin{tabular}{c|c|c|c|c}
    \toprule
    frequency [Hz] & $\lambda$ (deg) & $\beta$ (deg) & $\dot{f}$ [Hz/s] & $\rm{CR}$  \\ \hline
    61.4416 & 76.6500 & 60.5462 & $-4.6745\cdot 10^{-13}$ & 6.1300 \\
    104.1381 & 85.5924 &$-59.5761$ & $-5.8598 \times 10^{-13}$ & 4.5150 \\
    305.1118 & 23.2246 & 50.5265 & $-5.3121\times 10^{-12}$ & 4.3600 \\
    \hline
    \end{tabular}
  \caption{Average parameters of outliers passing the second follow-up stage (FU2).} 
    \label{tab:candidates}
\end{table*}
The first one, at $\sim 61.44$~Hz, has been discarded by looking at the corresponding refined peakmap, where a transient disturbance in H1 data, starting $\sim 20$ days before the end of the run and at the exact frequency of the candidate, can be identified, see Fig.~\ref{fig:cand_61_H}.
The other two have been subject to the third follow-up step (FU3), with an increase of the FFT duration of, respectively, a factor of 10 and 12 with respect to the initial value used in the main search. Neither of the outliers pass the significance veto and, moreover, the CR reduces for both outliers (in both detectors). Thus we finally discard them as not compatible with astrophysical signals.

\subsection{Follow-up results for outliers with $W > 1$}

For the original outliers with $W>1$, we apply the follow-up procedure described in Sec.~\ref{sec:fu_viterbi}. In total, 7816 outliers are identified with $W>1$ in the main search. After the first step of searching data from both LIGO detectors using the Viterbi method, only a small number of the original outliers (267 out of 7816) are identified above the Viterbi threshold (1\% false alarm probability in each sub-band searched). After the second step, in which the Viterbi searches are carried out in data from Hanford and Livingston detectors separately, we eliminate an additional 73 outliers, and the remaining 194 are subject to further manual scrutiny.

Finally, we carry out manual inspections for the remaining 194 outliers, clustered at six different frequencies. We find 178 of them clustered near 26.33~Hz are caused by a pulsar hardware injection in O3, and thus are not of astrophysical origin. Another 15 are associated with various instrumental disturbances. For the last outlier found at $\sim 190.82$~Hz, further investigations in every few days' of data find that it results from disturbances around that frequency at the Hanford detector in the last $\sim 20$ days.

\section*{Acknowledgments}
This material is based upon work supported by NSF’s LIGO Laboratory which is a major facility
fully funded by the National Science Foundation.
The authors also gratefully acknowledge the support of
the Science and Technology Facilities Council (STFC) of the
United Kingdom, the Max-Planck-Society (MPS), and the State of
Niedersachsen/Germany for support of the construction of Advanced LIGO 
and construction and operation of the GEO600 detector. 
Additional support for Advanced LIGO was provided by the Australian Research Council.
The authors gratefully acknowledge the Italian Istituto Nazionale di Fisica Nucleare (INFN),  
the French Centre National de la Recherche Scientifique (CNRS) and
the Netherlands Organization for Scientific Research (NWO), 
for the construction and operation of the Virgo detector
and the creation and support  of the EGO consortium. 
The authors also gratefully acknowledge research support from these agencies as well as by 
the Council of Scientific and Industrial Research of India, 
the Department of Science and Technology, India,
the Science \& Engineering Research Board (SERB), India,
the Ministry of Human Resource Development, India,
the Spanish Agencia Estatal de Investigaci\'on (AEI),
the Spanish Ministerio de Ciencia e Innovaci\'on and Ministerio de Universidades,
the Conselleria de Fons Europeus, Universitat i Cultura and the Direcci\'o General de Pol\'{\i}tica Universitaria i Recerca del Govern de les Illes Balears,
the Conselleria d'Innovaci\'o, Universitats, Ci\`encia i Societat Digital de la Generalitat Valenciana and
the CERCA Programme Generalitat de Catalunya, Spain,
the National Science Centre of Poland and the European Union – European Regional Development Fund; Foundation for Polish Science (FNP),
the Swiss National Science Foundation (SNSF),
the Russian Foundation for Basic Research, 
the Russian Science Foundation,
the European Commission,
the European Social Funds (ESF),
the European Regional Development Funds (ERDF),
the Royal Society, 
the Scottish Funding Council, 
the Scottish Universities Physics Alliance, 
the Hungarian Scientific Research Fund (OTKA),
the French Lyon Institute of Origins (LIO),
the Belgian Fonds de la Recherche Scientifique (FRS-FNRS), 
Actions de Recherche Concertées (ARC) and
Fonds Wetenschappelijk Onderzoek – Vlaanderen (FWO), Belgium,
the Paris \^{I}le-de-France Region, 
the National Research, Development and Innovation Office Hungary (NKFIH), 
the National Research Foundation of Korea,
the Natural Science and Engineering Research Council Canada,
Canadian Foundation for Innovation (CFI),
the Brazilian Ministry of Science, Technology, and Innovations,
the International Center for Theoretical Physics South American Institute for Fundamental Research (ICTP-SAIFR), 
the Research Grants Council of Hong Kong,
the National Natural Science Foundation of China (NSFC),
the Leverhulme Trust, 
the Research Corporation, 
the Ministry of Science and Technology (MOST), Taiwan,
the United States Department of Energy,
and
the Kavli Foundation.
The authors gratefully acknowledge the support of the NSF, STFC, INFN and CNRS for provision of computational resources.

This work was supported by MEXT, JSPS Leading-edge Research Infrastructure Program, JSPS Grant-in-Aid for Specially Promoted Research 26000005, JSPS Grant-in-Aid for Scientific Research on Innovative Areas 2905: JP17H06358, JP17H06361 and JP17H06364, JSPS Core-to-Core Program A. Advanced Research Networks, JSPS Grant-in-Aid for Scientific Research (S) 17H06133 and 20H05639 , JSPS Grant-in-Aid for Transformative Research Areas (A) 20A203: JP20H05854, the joint research program of the Institute for Cosmic Ray Research, University of Tokyo, National Research Foundation (NRF) and Computing Infrastructure Project of KISTI-GSDC in Korea, Academia Sinica (AS), AS Grid Center (ASGC) and the Ministry of Science and Technology (MoST) in Taiwan under grants including AS-CDA-105-M06, Advanced Technology Center (ATC) of NAOJ, Mechanical Engineering Center of KEK.


{\it We would like to thank all of the essential workers who put their health at risk during the COVID-19 pandemic, without whom we would not have been able to complete this work.}

\bibliographystyle{ieeetr}
\bibliography{references}

\iftoggle{endauthorlist}{
  \let\author\myauthor
  \let\affiliation\myaffiliation
  \let\maketitle\mymaketitle
  \title{The LIGO Scientific Collaboration, Virgo Collaboration, and KAGRA Collaboration  }
  \pacs{}


\author{R.~Abbott}
\affiliation{LIGO Laboratory, California Institute of Technology, Pasadena, CA 91125, USA}
\author{H.~Abe}
\affiliation{Graduate School of Science, Tokyo Institute of Technology, Meguro-ku, Tokyo 152-8551, Japan  }
\author{F.~Acernese}
\affiliation{Dipartimento di Farmacia, Universit\`a di Salerno, I-84084 Fisciano, Salerno, Italy  }
\affiliation{INFN, Sezione di Napoli, Complesso Universitario di Monte S. Angelo, I-80126 Napoli, Italy  }
\author{K.~Ackley\,\orcidlink{0000-0002-8648-0767}}
\affiliation{OzGrav, School of Physics \& Astronomy, Monash University, Clayton 3800, Victoria, Australia}
\author{N.~Adhikari\,\orcidlink{0000-0002-4559-8427}}
\affiliation{University of Wisconsin-Milwaukee, Milwaukee, WI 53201, USA}
\author{R.~X.~Adhikari\,\orcidlink{0000-0002-5731-5076}}
\affiliation{LIGO Laboratory, California Institute of Technology, Pasadena, CA 91125, USA}
\author{V.~K.~Adkins}
\affiliation{Louisiana State University, Baton Rouge, LA 70803, USA}
\author{V.~B.~Adya}
\affiliation{OzGrav, Australian National University, Canberra, Australian Capital Territory 0200, Australia}
\author{C.~Affeldt}
\affiliation{Max Planck Institute for Gravitational Physics (Albert Einstein Institute), D-30167 Hannover, Germany}
\affiliation{Leibniz Universit\"at Hannover, D-30167 Hannover, Germany}
\author{D.~Agarwal}
\affiliation{Inter-University Centre for Astronomy and Astrophysics, Pune 411007, India}
% ...

\collaboration{The LIGO Scientific Collaboration, the Virgo Collaboration, and the KAGRA Collaboration}



  \newpage
  \maketitle
}

\end{document}